\documentclass[%
 reprint,
 amsmath,amssymb,
 aps,
prl,
]{revtex4-2}

\usepackage{graphicx}
\usepackage{dcolumn}
\usepackage{bm}
\usepackage{xcolor}
\usepackage[normalem]{ulem}

\usepackage{comment}
\usepackage{xr}
\usepackage{natbib}
\begin{document}

\preprint{APS/123-QED}

\title{Learning to shine: Neuroevolution enables optical control of phase transitions}

\author{Sraddha Agrawal}
\email{sagrawal@anl.gov}
\affiliation{Center for Nanoscale Materials, Argonne National Laboratory, Lemont, IL 60439, USA}

\author{Stephen Whitelam}
\affiliation{Molecular Foundry, Lawrence Berkeley National Laboratory, Berkeley, CA 94720, USA}

\author{Pierre Darancet}
\email{pdarancet@anl.gov}
\affiliation{Center for Nanoscale Materials, Argonne National Laboratory, Lemont, IL 60439, USA}
 \affiliation{Computational Science Division, Argonne National Laboratory, Lemont, IL 60439, USA}
 \affiliation{Northwestern Argonne Institute of Science and Engineering, Evanston, IL 60208, USA}

\date{\today}

\begin{abstract}

We address active optical steering of structural phase transitions in solids. We demonstrate that existing evolutionary reinforcement learning approaches can derive optimal time-dependent electric fields in driven dissipative classical systems far beyond the harmonic regime, enabling the stabilization of non-thermal structural phases. Our approach relies on experimentally-extractable metrics of the phase-space evolution and interpretable Fourier Neural Network surrogates of the electric field. 
Using this method on first-principles models, we obtain protocols stabilizing a symmetric phase in bismuth through impulsive Raman and displacive excitations with continuous and pulsed light sources in the presence of dissipation and thermal disorder. The method is gradient-free, enabling optimization loops based solely on experimental data and providing a practical route for controlling light-induced structural dynamics independently of the microscopic model of light-matter interactions.
\end{abstract}

\maketitle

\section{Introduction}
The structural evolution of materials under electric fields has long been a focus of study in various fields, including coherers for telegraphy~\cite{branly1890variations,bose1901change,dilhac2009edouard}, material reliability in CMOS architectures~\cite{islam2007recent}, and non-Turing computing architectures~\cite{di2022memcomputing}. For time-varying electric fields, from the THz to visible frequency range, a technologically significant application lies in the transient stabilization of structural phases of high dielectric contrast compared to their ground state~\cite{forst2011nonlinear}. Despite experimental successes~\cite{zeng2025photo,forst2011nonlinear,shukla2026metastable}, discoveries of such phases have relied on serendipitous approaches to estimate light-induced transition pathways from the linear response properties of materials.  As quantitative models of the interaction between structural degrees of freedom and time-varying electric fields are becoming available for various classes of materials~\cite{kaganov1957relaxation,laubereau1978vibrational,allen1987theory,zeiger1992theory,dhar1994time,merlin1997generating,forst2011nonlinear,Yan_PNAS,Calandra_PRL,sadasivam2017theory,juraschek2018sum,Benedek_PRX,Benedek_PRB,Yan_PRB,Calandra_PRB,Abajo_Science}, a method to derive optimal illumination protocols that can guide material systems into putative non-equilibrium states through highly anharmonic potential energy surfaces~\cite{Haldar24} is increasingly desirable.

In this work, we propose an evolutionary reinforcement learning approach based on neuroevolution, optimizing phase-space trajectories that can derive optimal time-dependent electric fields that stabilize non-equilibrium structural phases. 
Our approach minimizes metrics extracted from the phase-space evolution to optimize Fourier Neural Network (FNN) surrogates of the classical AC electric field. 
By applying this method to different models of the light-induced phase transition in bulk bismuth, we demonstrate optimization far beyond the harmonic regime for both continuous-wave and pulsed laser protocols in the presence of thermal noise and dissipation. Our FNN surrogate is interpretable and allows for experimental constraints on laser power, repetition rates, and pulse shape. We also show that the protocols can be evaluated and optimized solely based on experimental data, e.g. from the half-periods in transient reflectivity oscillations in broken-symmetry systems. Our framework therefore provides a general platform for control over nonlinear lattice dynamics, robustness against thermal fluctuations, and adaptive optical control strategies in complex quantum materials.

In the next, our objective is to optimize a classical time-varying electric field $\mathbf{E}(t)$  to drive the system from a thermally equilibrated, light-free configuration defined by internal coordinates and velocities $ \{ \mathbf{X} , \dot{\mathbf{X}} \} (t=0) $ to a user-defined set of coordinates $ \{ \mathbf{X}_{\textrm{target}} , \dot{\mathbf{X}}_{\textrm{target}} \} $ over a finite time $t \in \left[0; t_\textrm{target}\right] $. We consider the structural evolution of a system under the Born-Oppenheimer approximation, obeying a Langevin-like equation:
\begin{equation}
m_i \ddot{\mathbf{X_i}} = - \left. \frac{d U  }{d \mathbf{X_i}}\right|_{\mathbf{X}(t)} - \gamma \dot{\mathbf{X}_i}   + \mathbf{F}_{\mathrm{R}}(t) +  \int_{0}^t d\tau \mathcal{K} \left(\mathbf{E}(\tau),\mathbf{X}(\tau) \right).
\label{Eq:EoM}
\end{equation}
The first, second and third terms on the right hand side represent the  force along \( \mathbf{X_i} \) from the light-free potential energy surface (PES) \( U(\mathbf{X}) \), the friction force  where $\gamma$ is the damping coefficient and the random force, respectively.  The fourth term accounts for the nonlinear and memory-dependent coupling with the classical electromagnetic field, the functional form of  $\mathcal{K} \left(.,. \right)$ depends on the microscopic coupling mechanism. 

\begin{figure*}[htbp!]
    \centering
    \includegraphics[width=\textwidth]{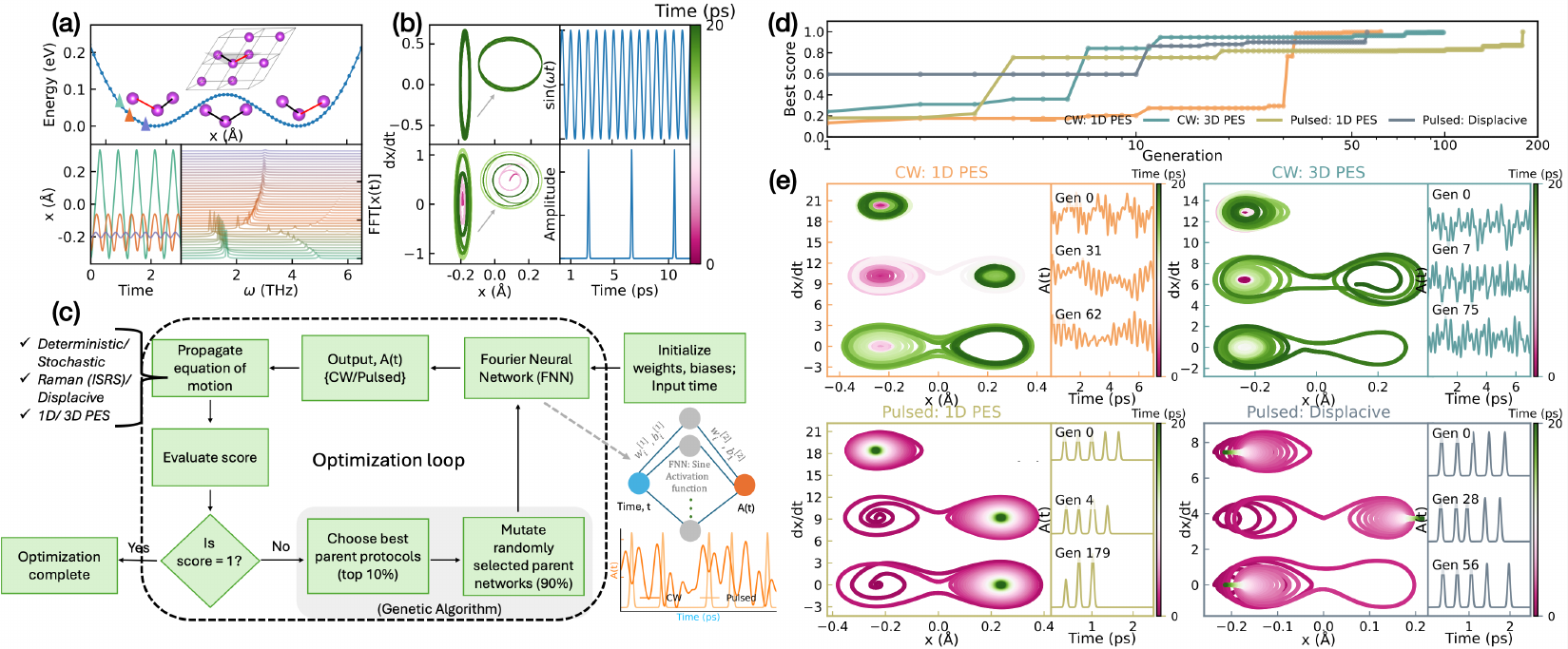}
    \caption{ (a) Potential energy surface along the A$_{1g}$ displacement vector (top), Time evolution of phonon amplitude starting from three different initial conditions (bottom left), and waterfall plot of the Fourier transforms of the oscillations from small (top) to large (bottom) amplitude oscillations. (b) Phase portraits of the system evolution under sinusoidal (top) and pulsed (bottom) harmonic protocols. Time coloring denotes trajectory evolution. (c) Optimization framework for the FNN (amplitude function for both CW and pulsed protocols) for different PES dimensions, and microscopic models of light-matter interaction. (d) Best score versus generation for CW and pulsed driving on 1D and 3D PESs, and displacive excitation dynamics. (e) Associated representative phase-space trajectories  and corresponding protocols A(t) at different stages of the optimization.}
    \label{fig:harmonic_protocol} 
\end{figure*}
 
To test the generality of our approach, we optimize the electric field for two distinct microscopic coupling mechanisms. We first consider a configuration-dependent, Markovian impulsive Raman scattering term~\cite{laubereau1978vibrational,dhar1994time,merlin1997generating,forst2011nonlinear, juraschek2018sum,Haldar24} proportional to the square of the instantaneous field, i.e. $\mathcal{K} \left(\mathbf{E}(\tau),\mathbf{X}(\tau) \right) = \mathbf{E}^*(\omega, \tau) \left. \frac{d \chi(\omega)}{d\mathbf{X_i}}\right|_{\mathbf{X}(\tau)} \mathbf{E}(\omega, \tau) \delta(\tau -t )$. \( \mathbf{E}(\omega, t) = \mathbf{A}(t) \ e^{i\omega t} \) is the electric field and \( \mathbf{A}(t) \) is the slowly-varying amplitude representing the pulse-time evolution, and \( e^{i\omega t} \) is the rapidly oscillating monochromatic component. The optomechanical coupling coefficient for degree of freedom $\mathbf{X_i}$ is the configuration-dependent Raman cross-section $ R_{\mathbf{X}} (\omega; \mathbf{X}_i) = \left. \frac{d \chi(\omega)}{d\mathbf{X_i}}\right|_{\mathbf{X}(t)} $, where \( \chi (\omega)\) is the macroscopic polarizability tensor of the system at frequency $\omega$ in configuration ${\mathbf{X}(t)}$. Due to the separation of timescales between the fast oscillations and the slow variation of \( \mathbf{A}(t) \) and the structural evolution, the external force \( \approx | \mathbf{A}(t)|^2 \left. \frac{d \chi(\omega)}{d\mathbf{X_i}}\right|_{\mathbf{X}(t)}\) for an isotropic medium. Alternatively, we also consider a displacive model of light-matter interactions~\cite{zeiger1992theory,caruso2023quantum,emeis2025coherent} leading to an additional, non-Markovian term depending on the field and configuration at past times through the instantaneous density of excited carriers. Equations and models used in this work and their parameterizations using density functional theory (DFT) and time-dependent density functional theory (TDDFT) are described in the Methods section. Convergence tests and additional analyses are provided in the Supplementary Information (SI).

\section{Results and discussion}
 Figure~\ref{fig:harmonic_protocol}(a) shows the computed 1-dimensional PES \(U(x)\) along the fully symmetric structural A\(_{1g}\) coordinate of bulk bismuth obtained from DFT~\cite{Haldar24} (PES along the $E_g$ coordinates can be found in Fig. S8). Bismuth atoms form a bipartite rhombohedral network, with two equivalent atoms, Bi$_{A}$ and Bi$_{B}$, per unit cell placed on the long diagonal of the rhombohedron (the \( x \) coordinate corresponding to the A\(_{1g}\) optical phonon, in Figure \ref{fig:harmonic_protocol}(a) is derived from the Bi$_{A}$-Bi$_{B}$ distance along that diagonal). We refer to \( x = 0 \) as the structure in which the intra-unit cell distance Bi$_{A}$-Bi$_{B}$ is half of the diagonal of the rhombohedron. In this high-symmetry configuration, all $A/B$ atoms are located at the center of a regular octahedron of $B/A$ atoms, with an energy of 0.1 eV/unit cell above the minima at \( x = \pm 0.25 \)Å. In the two-fold degenerate, Peierls-like structural ground-state,  each Bi$_{A/B}$ is located slightly off-center from the regular Bi$_{B/A}$ octahedron, with the 3 nearest atoms located at 3.10 Å and the 3 next-nearest at 3.49 Å. While not thermodynamically stable, the \( x = 0 \) configuration has a more metallic character (Fig.S7 (a)) than the broken symmetry ground-state, resulting in \( x = 0 \) being both an energy minimum at very high excited carrier density~\cite{zeiger1992theory}, and a zero of the Raman coupling at all optical frequencies~\cite{Haldar24} (as can be seen in Figure S7). This implies that the external force term in Eq.~\ref{Eq:EoM} for both displacive and Raman processes is pushing the structure towards \( x = 0 \), yet with distinct configuration- and time-dependencies~\cite{zeiger1992theory}. Also shown in Figure~\ref{fig:harmonic_protocol}(a)  are the computed undamped oscillations for different initial amplitudes using the first term of Eq.~\ref{Eq:EoM}, and their Fourier transforms. At small amplitudes, the system exhibits sinusoidal oscillations at $\simeq$ 3 THz, in good agreement with experiment~\cite{Bi_2}, with significant phonon softening ($>$10\% of the harmonic frequency) and higher harmonics appearing as the apex of the orbits approaches the saddle point. At larger amplitude, the system experiences bifurcation, with a strongly non-sinusoidal oscillation comprised of two frequencies $\simeq$ 1.2 THz and 3.6 THz which harden with increasing amplitude. 
The amplitude dependence of both the optomechanical cross-section and oscillatory dynamics has detrimental effects on the ability of harmonic illumination protocols (i.e. $A(t) \propto \textrm{sin}(\omega_{ph} t/n)$ with $n$ an integer) to drive the system across its phase transition, as shown in Figures~\ref{fig:harmonic_protocol}(b) for continuous wave and pulsed protocols with constant time delays corresponding to integer multiples of the phonon periods. As shown in the corresponding phase portraits, the resulting orbits are confined to one side of the \( \dot{x} \) axis and fail to achieve a phase transition, with the efficiency of the harmonic drive vanishing as the system approaches the saddle point~\cite{Haldar24}.

We will now show that the neuroevolution framework proposed in refs~\cite{Whitelam2020, Whitelam2023_2, Whitelam2023}, where neural networks are trained through evolutionary reinforcement learning to control self-assembly processes, can be adapted to the problem of optical phase transitions by defining metrics based on phase-space evolution: Figure~\ref{fig:harmonic_protocol}(c) summarizes the optimization workflow using an FNN~\cite{ngom2021fourier} with time as input to represent A(t). Concretely, the output of each hidden layer node $i$ is $\sin\left(w_i^{[1]} t + b_i^{[1]}\right)$ with the weights \(w_i\) and biases \(b_i\) physically associated with an angular frequency and phase. As discussed in Section~\ref{sec:fnn} in the Methods, depending on the continuous or pulsed nature of the protocols under investigation, the same architecture can be used with a different expression of the pulse form. FNNs have shown promise in solving stochastic partial differential equations with oscillatory dynamics~\cite{wang2021eigenvector} and have obvious expressive power that leads to compact representations. After initialization of the weights (either random or as pairs of frequencies differing by the harmonic phonon frequency), a genetic algorithm optimizes the FNN by modulating its weights and biases, based on a score computed from the system's evolution simulated using Eq.~\ref{Eq:EoM}. We quantify the success of a protocol A(t) using a score function on individual phase-space orbits of the trajectories.  To this end, the total simulation time \( T \)  is divided into time periods based on the crossings at times \( t_i \) and \( t_{i+1} \) of the velocity \( \dot{x} = 0 \) axis. This approach is akin to a Poincaré surface of sections and is generalizable to any dimension of the PES (as shown below).
Taking the expectation value of the position coordinate over the \(i\)-th orbit as:
\begin{equation}
    \langle x \rangle_i =
    \frac{1}{\Delta T_i}
    \int_{t_i}^{t_{i+1}} x_i(t)\,dt,
\end{equation}
We define the fitness of a protocol as 
\begin{equation}
\text{Score} = \max_{i\in \textrm{traj}}\left( 1 - \left| \frac{\langle x \rangle_i -x_{\text{saddle}} }{x_{\textrm{min}}-x_\text{saddle}} \right| \right) ,    
\end{equation}
where \(x_{\text{min}}\approx 0.25\,\text{\AA}\) and \(x_{\text{saddle}}=0\) are the ground- and saddle state configurations.  This rewards trajectories by evaluating their ``best'' orbit, with a score of 1 corresponding to a protocol leading to at least one orbit of average position \( \langle x \rangle \) aligned with the saddle point, and lower scores corresponding to orbits around the ground state. As shown in the following, the same optimization metrics are applicable to dynamical systems governed by stochastic differential equations: For thermal distributions, the score is evaluated on an ensemble of trajectories and becomes a distribution function.  

The convergence of the genetic algorithm and the impact of different hyperparameters are discussed in Sec. III of the Supplementary Information. Briefly, we find that convergence is accelerated by reintroducing the top 10\% of the parent networks in the next generation, while the remaining 90\% of the population is generated by mutating the top parent networks, using a mutation strength of 0.2.

The results of the optimization are shown in Figure~\ref{fig:harmonic_protocol}(d) for a variety of microscopic mechanisms (Raman/displacive), pulse shapes (CW/Pulsed) and dimensionality of the PES (1D/3D).  In all cases, optimization progressively discovers higher-scoring control protocols, eventually approaching scores close to unity. Although convergence behavior depends on the excitation mechanism and dimensionality of the underlying energy landscape, efficient protocols are consistently identified within a small number of generations. Representative optimized phase-space trajectories and their corresponding control fields are shown in Figures~\ref{fig:harmonic_protocol}(e). The optimized protocols differ substantially across excitation mechanisms and PES dimensionalities, indicating that the learning framework adapts to the underlying nonlinear dynamics rather than converging to a universal harmonic resonance condition. Similarly, protocols across physical mechanisms generate qualitatively different dynamical pathways. In particular, pulsed excitation protocols exploit the increase in Raman cross-section away from the saddle point, whereas their counterpart of displacive excitations modifies the effective force acting on the structural coordinate and produces distinct phase-space trajectories. Such learned control strategies are easily interpreted by looking at the Fourier decompositions of the optimized CW protocols, presented in Figure S2, along with protocol efficiency and pulse-timing analyses for pulsed driving in Figures S18 - S20. Such analysis reveals that the optimization does not simply converge to a single resonant driving frequency, but instead exploits combinations of frequencies and pulse-timing patterns adapted to the nonlinear structural dynamics.

\begin{figure*}[htbp!]
    \centering
    \includegraphics[width=\textwidth]{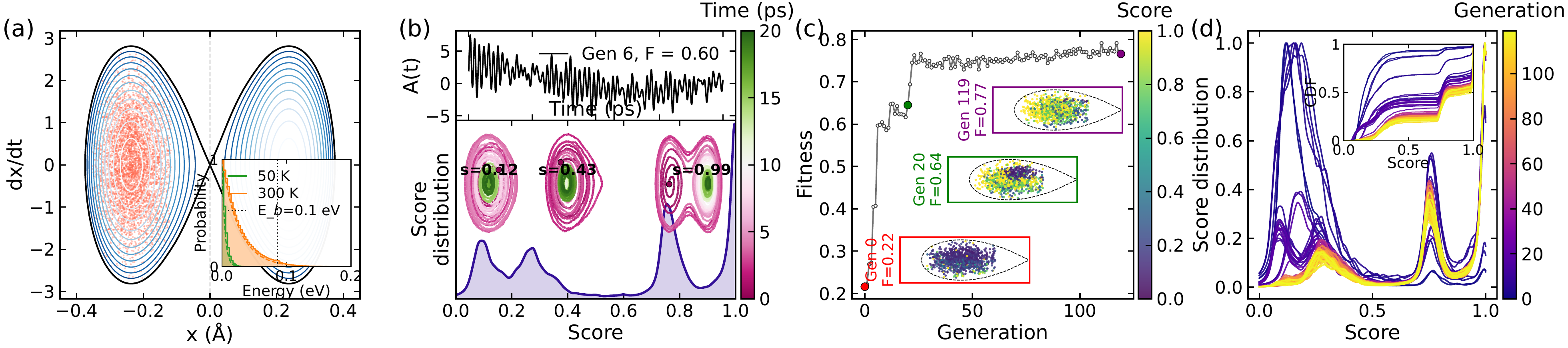}
    \caption{Optimization in the presence of thermal distributions and noise. (a) Langevin thermal sampling without external driving, showing canonical probability distributions at 50 K and 300 K on the anharmonic double-well PES relative to the barrier energy. Inset: corresponding energy probability distributions. (b) Protocol (top) from generation 6 and associated distribution $\mathcal{D}(s)$ of scores for $1000$ initial conditions, with examples of low-, intermediate-, and high-scoring phase-space trajectories. (c) Evolution of the best fitness during optimization at 50 K, with score vs initial phase space configuration in inset. (d) Evolution of the score distributions for the best protocol from each generation  (inset: associated CDFs).}
    \label{fig:langevin_protocol}
\end{figure*}

We  now examine the optimization of protocols in the presence of stochastic effects and thermal disorder. Setting 
$\mathbf{F}_{\rm R}
=
\sqrt{2m_i\gamma k_B T}\,
\boldsymbol{\eta}_i(t)$,
where $\boldsymbol{\eta}_i(t)$ is Gaussian white noise satisfying
$\langle\boldsymbol{\eta}_i(t)\rangle=0$ and
$\langle\eta_{i,\alpha}(t)\eta_{j,\beta}(t')\rangle
=
\delta_{ij}\delta_{\alpha\beta}\delta(t-t')$ in Eq.~\ref{Eq:EoM} and operating from a thermal distribution of initial configurations then makes the score function a distribution of scores of individual trajectories. Figure~\ref{fig:langevin_protocol}(a) shows the canonical probability distributions obtained from the Langevin dynamics on the Bi PES in the absence of any driving field. We perform optimization for impulsive Raman scattering for ensembles at 50 K, which ensures that successful barrier-crossing dynamics results only from the learned  protocols. Figure~\ref{fig:langevin_protocol}(b) shows the score distribution $\mathcal{D}(s)$ generated by a CW protocol applied to 1000 initial configurations drawn from the thermal distributions, along with representative low-, intermediate- and high-scoring phase portraits of three initial conditions. The fitness of a protocol is simply derived from the distance of the score distribution from the ideal Heaviside function $H(x-1)$. In practice, we find that optimization is facilitated by a fitness definition balancing a CDF metric $\left(
\int_0^1
\left[
\mathcal{D}(s)-H(s-1)
\right]^2
ds
\right)^{1/2}$(to ensure the maximum number of initial configurations results in a trajectory crossing the saddle point) with a score-consistency  metric $
\left(
\frac{1}{N}
\sum_{i=1}^{N}
(1-s_i)^4
\right)^{1/4}$ of the score $s$ across N initial conditions.  For each protocol, \(N=1000\) trajectories are evaluated using initial conditions randomly sampled from a pool of 5000 thermalized configurations.

The result of the  optimization is shown in Figure~\ref{fig:langevin_protocol}(c). Representative ensembles from generations 0, 24, and 117 illustrate the progressive emergence of high-scoring dynamical responses across the thermally sampled initial conditions. As optimization proceeds, an increasing fraction of trajectories approach the ideal score of unity, indicating that protocols become robust across a broader range of initial conditions rather than a small subset of exceptional realizations.  Figure~\ref{fig:langevin_protocol}(d) shows the corresponding evolution of the distributions together with their CDF (inset) showing robust learning across initial conditions and in the presence of thermal noise.

\begin{figure}[htbp]
    \centering
    \includegraphics[width=\columnwidth]{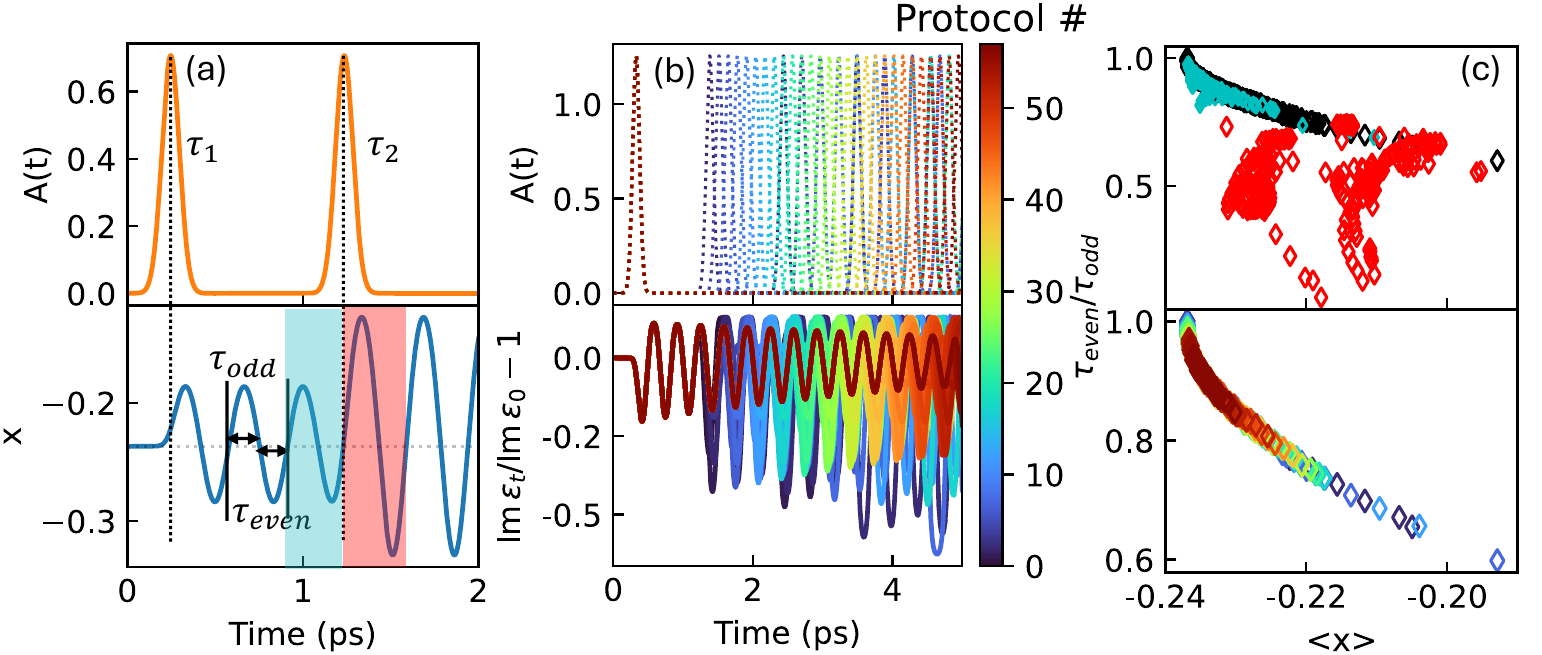}
    \caption{ (a) Illustration of a two-pulse protocol (top panel) and the corresponding time evolution of position (bottom), (b) Harmonic pulse driven protocols with varying time delays (top) and the corresponding normalized imaginary part of the dielectric function with time (bottom), (c) Correlation of the ratio (as defined in (a)) with the average position for all harmonic protocols; eliminating the data in red and blue (refers to the points in the time window around the pulse timing as marked in (a)) gives rise to linear relationship between the two quantities (bottom panel).}
    \label{fig:expt_correlation}
\end{figure}

We conclude by discussing the implementation of our framework for self-driven experiments. As our optimization framework is gradient-free, automatically determining an experimental score for a given protocol suffices to replace the simulation of the evolution in the optimization loop. Figure~\ref{fig:expt_correlation} (a) shows the simulated evolution of x(t) in a 2-pulse protocol, with two distinct pulses at times $\tau_1 \text{ and } \tau_2$. Half-periods of evolution $\tau_{even} \text{ and } \tau_{odd}$, are defined within each oscillation period. The symmetry of the PES imposes that $\tau_{even} \neq \tau_{odd}$ as the potential becomes softer near $x=0$. Consequently, strongly asymmetric oscillations of the imaginary part of the dielectric function $\mathrm{Im}(\epsilon)$ occur along the trajectories of all protocols simulated in Figure~\ref{fig:expt_correlation} (b), with each trajectory yielding a set of $\tau_{even}/\tau_{odd}$. As shown in  Figure~\ref{fig:expt_correlation} (c), the asymmetry of the oscillations $\tau_{even}/\tau_{odd}$ is a monotonic function of the order parameter and therefore can be used as a scoring function of a protocol (excluding the periods synchronous to a pulse hitting the system), enabling direct optimization (an example experimental framework is shown in Figure S21). 

In conclusion, we have developed a neuroevolution based reinforcement learning framework capable of optimizing time-dependent electric field amplitudes to control optically induced phase transitions in quantum materials. By applying this framework to a broken symmetry material, bismuth (Bi), we have demonstrated that Fourier Neural Networks can learn and generate optimized illumination protocols for phonon amplification near phase transitions, extending control beyond the harmonic regime in noisy conditions and in the presence of thermal disorder. Our approach introduces a systematic method for identifying effective and interpretable protocols, replacing trial-and-error methods, and can be applied to optimize non-equilibrium phase transitions across a range of nonlinear systems.  Furthermore, we showed that our method can, in principle, be applied to entirely experimental setups without the need for theoretical input for the stabilization of high-symmetry phases of broken-symmetry materials.

\section*{Methods}

\subsection*{First-principles calculations}

 The parameters in Eq. \ref{Eq:EoM} in the main text were computed for the case of bulk Bismuth using the Vienna \emph{Ab initio} Simulation Package (VASP) 6.3.1.\cite{vasp1,vasp2,vasp3}. We carried out first-principles density functional theory (DFT) and linear response time-dependent DFT calculations on the primitive unit cell of Bi, consisting of two atoms. 
The exchange–correlation energy was calculated with the generalized gradient approximation of Perdew–Burke–Ernzerhof (PBE) functional and ultrasoft pseudopotentials.\cite{vasp_paw,vasp_pp} The plane-wave energy cutoff was set at 350 eV and a k-point grid of 18x18x18 was used in conjunction with the tetrahedron method to obtain a smooth dielectric function. A total of 128 bands (113 empty bands) and Methfessel-Paxton method of order 2 with 0.2 eV smearing were included in the calculation. The convergence of dielectric function with respect to the number of k-points and the number of bands is shown in Figures S3 - S6. The imaginary plasma frequency was set to 25.8 eV to account for the intraband transitions in the dielectric function, and local-field effects were included within the Random-Phase approximation in the frequency-dependent dielectric function.
Phonopy was used to compute the zone-centered phonon eigenvectors and frequencies\cite{phonopy1, phonopy2}. The potential energy was evaluated using finite displacements in the atomic simulation environment~\cite{ase} and interpolated using a spline function in Scipy\cite{scipy}. The dielectric function was fitted to a polynomial regression model using only even powers up to the 8th degree. The derivatives of all quantities were evaluated as derivatives of the splines.

\subsection*{Lattice dynamical models}

\paragraph{\textbf{Model I: One-dimensional impulsive Raman dynamics.}}

For the one-dimensional simulations, the structural coordinate $q$ corresponds to the fully symmetric $A_{1g}$ phonon amplitude. The lattice dynamics are governed by
\begin{equation}
m\ddot q=
-\frac{\partial U(q)}{\partial q}
-\gamma\dot q
-\alpha R(q)A^2(t),
\label{eq:1d_raman}
\end{equation}

where $U(q)$ is the first-principles potential energy surface, $R(q)=\partial\chi/\partial q$ is the configuration-dependent Raman coupling, $\alpha$ determines the overall field strength, and $A(t)$ is the slowly varying electric-field envelope generated by the optimization algorithm. The Raman force is proportional to the instantaneous optical intensity and continuously evolves with the lattice configuration through the Raman tensor derivative. A phenomenological damping rate of 0.1~THz is included in all deterministic simulations.

\paragraph{\textbf{Model II: Three-dimensional impulsive Raman dynamics.}}

To investigate multidimensional lattice dynamics, the optimization was extended to the three-dimensional structural coordinate
\begin{equation}
\mathbf{Q}=(Q_{A_{1g}},Q_{E_g^1},Q_{E_g^2}),
\end{equation}

where the doubly degenerate $E_g$ modes are coupled to the fully symmetric $A_{1g}$ coordinate through the anharmonic potential energy surface. The equations of motion become
\begin{equation}
m\ddot{\mathbf Q}
=
-\nabla_{\mathbf Q}U(\mathbf Q)
-\gamma\dot{\mathbf Q}
-
\alpha
\begin{pmatrix}
R(Q_{A_{1g}})\\
0\\
0
\end{pmatrix}
A^2(t).
\label{eq:3d_raman}
\end{equation}

Only the $A_{1g}$ coordinate couples directly to the optical field through impulsive Raman scattering, whereas the $E_g$ coordinates evolve through their anharmonic coupling within the multidimensional potential energy surface. This formulation enables the optimization framework to capture nonlinear mode coupling beyond the single-mode approximation.

\paragraph{\textbf{Model III: Displacive excitation dynamics.}}

To model displacive excitation of coherent phonons, the lattice dynamics are coupled to the evolution of the nonequilibrium photoexcited carrier population. The governing equations are

\begin{align}
\dot x &= v,\\
\dot v &= -\frac{1}{m}\left(\frac{dU(x)}{dx}+2c^2nx\right)-\gamma v,\\
\dot n &= \Gamma(x)\alpha A^2(t)-\beta n,
\end{align}

where $x$ is the structural coordinate, $v$ is the lattice velocity, and $n$ is the nonequilibrium carrier population generated by optical excitation. The parameter $\beta$ represents carrier relaxation, whereas $c$ controls the carrier-induced modification of the effective potential energy surface. The excitation rate is given by

\begin{equation}
\Gamma(x)=2f_{\rm excited}n^2(x)\left[1-R_{\rm opt}(x)\right],
\end{equation}

where the excited-state density and optical reflectivity are both obtained from first-principles calculations. The configuration dependence of these quantities allows the optical excitation rate to evolve self-consistently during the structural dynamics. The parameters employed for bismuth follow previous first-principles studies, while the optical driving field $A(t)$ is optimized using the same Fourier Neural Network representation and evolutionary optimization procedure employed for the impulsive Raman simulations.

\paragraph{\textbf{Model IV: Finite-temperature lattice dynamics.}}

Finite-temperature simulations were performed using the stochastic extension of the one-dimensional impulsive Raman model introduced in the main text. The system was first thermalized in the absence of external driving to generate a canonical distribution of initial positions and velocities. For each candidate optical protocol, $N=1000$ trajectories were propagated from initial conditions randomly sampled from a pool of 5000 thermalized configurations. The resulting trajectory scores were treated as an ensemble distribution and used to evaluate the finite-temperature fitness described in the Results.

\subsection*{Fourier neural network representation of optical protocols}
\label{sec:fnn}

We employ a unified Fourier Neural Network (FNN) architecture for both continuous-wave (CW) and pulsed driving protocols, with the primary difference arising from how the network outputs are interpreted. For CW excitation, the time-dependent field amplitude is generated explicitly by the network as a weighted sum of sinusoidal components,

\begin{equation}
A(t)=
\sum_{i=1}^{N_h}
w_i^{[2]}
\sin\!\left(
w_i^{[1]}t+b_i^{[1]}
\right)
+b^{[2]},
\label{eq:fnn}
\end{equation}

where $N_h$ is the number of hidden nodes. The first-layer weights $w_i^{[1]}$ and biases $b_i^{[1]}$ represent the angular frequencies and phases of the individual Fourier components, respectively, while the second-layer weights $w_i^{[2]}$ determine their amplitudes and the output bias $b^{[2]}$ determines the overall background. The first-layer biases are initialized from a Gaussian distribution with mean zero and standard deviation one, the second-layer weights are initialized from a Gaussian distribution with mean 0.8 and standard deviation one, and the output bias is fixed to zero. The number of hidden nodes can be restricted to retain an interpretable and experimentally accessible Fourier representation; the CW calculations discussed here use $N_h=10$ hidden nodes, each corresponding to a distinct tunable Fourier component.

The first-layer weights determining the Fourier frequencies are initialized either randomly or using pairs of frequencies separated by the harmonic phonon frequency, depending on the initialization scheme considered. During evolutionary optimization, the network parameters are perturbed through Gaussian mutations and constrained to the physically relevant parameter ranges. The resulting representation therefore allows the frequencies, phases, and relative amplitudes of the constituent Fourier components to evolve directly during optimization without requiring gradient-based training. Further analysis of the optimized frequency components, including the reduction of the ten-component CW protocol to a smaller subset while retaining comparable performance, is provided in Sec. I of the Supplementary Information.

For pulsed driving, the network does not generate the time-domain waveform explicitly. However, analogous to the Fourier components in the CW output, the network parameters define the arguments of a Gaussian pulse train in frequency space, the mathematical equation is given in Fig. S17. The pulsed case first constructs a frequency comb by combining Gaussian spectra at NN-determined positions, and then the subsequent inverse Fourier transform recovers the time-domain pulse sequence. The hidden-layer weights are initialized from the effective timescales and mapped to frequencies. These weights are subsequently constrained within physically relevant bounds and perturbed during optimization by Gaussian noise. The first-layer biases are fixed to a constant value, while the hidden–output connection is reduced to a constant scaling factor. The resulting set of parameters specifies the timing, carrier frequency, and envelope width of Gaussian pulses. The complete driving field is then constructed by summing over these Gaussian-modulated components, as shown by an example in Fig. S17 in the SI.

For each candidate network, the resulting CW or pulsed field $A(t)$ is applied to the corresponding lattice dynamical model. The structural trajectory generated by the equations of motion is assigned a score or ensemble fitness, and this quantity is used by the genetic algorithm to select and mutate the network parameters in the subsequent generation.

\section{Acknowledgments}
Work performed at the Center for Nanoscale Materials, a U.S. Department of Energy Office of Science User Facility, was supported by the U.S. DOE, Office of Basic Energy Sciences, under Contract No. DE-AC02-06CH11357. We would like to acknowledge computational resources from the Center for Nanoscale Materials HPC Cluster “Carbon”. SW performed work at the Molecular Foundry at Lawrence Berkeley National Lab supported by the Office of Science, Office of
Basic Energy Sciences, of the U.S. Department of Energy under Contract No. DE-AC02-05CH11231, and partly supported by US DOE Office of Science Scientific User Facilities AI/ML project “A digital twin for spatiotemporally resolved experiments”.

\section{Author contributions}

PD conceptualized the project and supervised the research. SW proposed the neuroevolution-based optimization approach and provided methodological guidance. SA developed and implemented the computational framework, performed all calculations and numerical analyses, and prepared the figures. All authors contributed to the interpretation of the results, refinement of the methodology, and preparation and revision of the manuscript.

\section{Competing interests}
All authors declare no financial or non-financial competing interests. 

\section{Data availability}

All data supporting the results reported in the main text are available at \url{https://github.com/Sraddha-Agrawal/L2S}. Large files and datasets corresponding to the Supplementary Information are not publicly available due to size constraints, but can be obtained from the authors upon reasonable request.

\section{Code availability}

The Python code implementing the optimization framework and analyses used in this study is publicly available at \url{https://github.com/Sraddha-Agrawal/L2S}.

\section{Supplementary Information}
The Supplementary Information contains additional analyses of the optimized Fourier components, hyperparameter sensitivity tests, dielectric-function convergence tests, potential-energy-surface projections, protocol-efficiency analyses, and details of the proposed experimental implementation.

\bibliography{biblio}

\end{document}


\preprint{APS/123-QED}

\title{Supplementary Information for Learning to shine: Neuroevolution enables optical control of phase transitions}

\author{Sraddha Agrawal}
\email{sagrawal@anl.gov}
\affiliation{Center for Nanoscale Materials, Argonne National Laboratory, Lemont, IL 60439, USA}

\author{Stephen Whitelam}
\affiliation{Molecular Foundry, Lawrence Berkeley National Laboratory, Berkeley, CA 94720, USA}

\author{Pierre Darancet}
\email{pdarancet@anl.gov}
\affiliation{Center for Nanoscale Materials, Argonne National Laboratory, Lemont, IL 60439, USA}
 \affiliation{Computational Science Division, Argonne National Laboratory, Lemont, IL 60439, USA}
 \affiliation{Northwestern Argonne Institute of Science and Engineering, Evanston, IL 60208, USA}
 
\date{\today}
             
\maketitle
\renewcommand{\thefigure}{S\arabic{figure}}
\renewcommand{\thetable}{S\arabic{table}}
\newpage

\onecolumngrid                

\setcounter{tocdepth}{2}      
\tableofcontents
\clearpage                    

\twocolumngrid

\section{Spectral analysis and reduction of optimized CW protocols}
\label{sec:frequency_reduction}

\begin{figure*}[htbp]
    \centering
    \includegraphics[width=\textwidth]{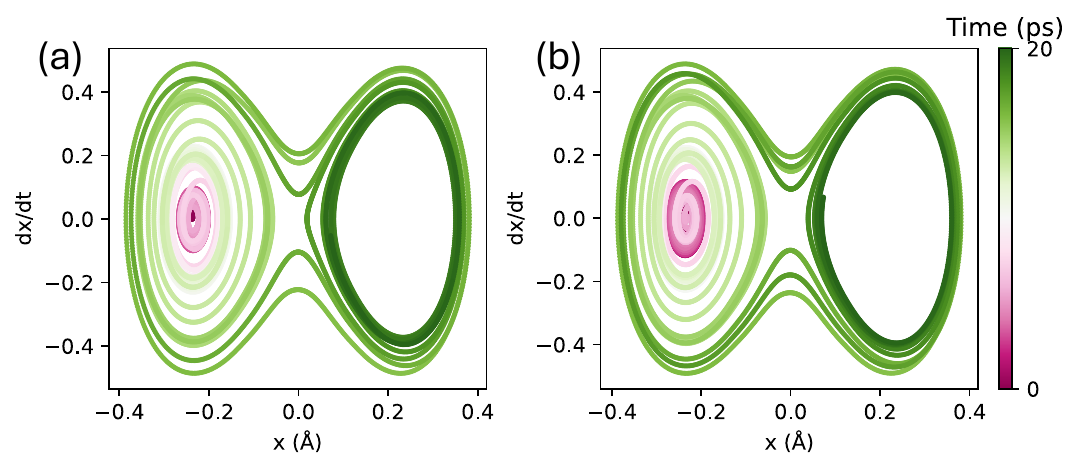}
    \caption{ (a) Full–basis CW optimized phase space using all ten sinusoidal components and, (b) Phase space corresponding to the five-component subset that attains essentially the same score as (a) }
    \label{fig:ten_to_five}
    \phantomsection
    \addcontentsline{toc}{section}{Figure~\thefigure: Phase space comparison for full basis and subset of CW protocols components}
\end{figure*}

To reduce the number of sine terms in Eq.~\ref{eq:fnn} of the main text, different subsets of the optimized weights \(w_i^{[2]}\) are evaluated. Starting from the full set of 10 optimized weights, all non-empty subsets are tested by retaining a chosen subset of weights and setting the others to zero. This results in \(2^{10}-1 = 1023\) possible combinations. In this way, it can be verified whether a smaller number of sine components is sufficient to achieve the same score as the full model.
Applying the method to the optimized CW protocol of Fig.~\ref{fig:harmonic_protocol}e, the ten learned frequencies (in THz) used in panel (a) of Fig.~\ref{fig:ten_to_five} are
\begin{align*}
\mathbf{f}\,[\mathrm{THz}] \approx \{\,2.825,\,-0.0621,\,3.716,\,-0.0601,\,2.695,\\
\qquad 0.2415,\,2.103,\,-0.2904,\,4.378,\,1.466\,\}.
\end{align*}

\begin{figure*}[htbp]
    \centering
    \includegraphics[width=\textwidth]{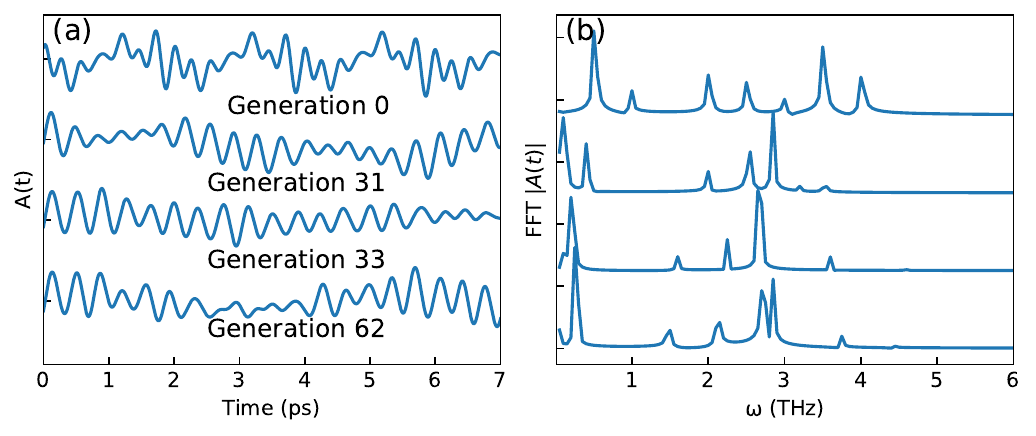}
    \caption{ Evolution of the optimized continuous-wave Raman-driving protocol on the one-dimensional potential energy surface corresponding to Fig. \ref{fig:harmonic_protocol}e of the main text. (a) Representative optical protocols from generations 0, 31, 33, and 62, (b)Fourier transforms of the corresponding protocols.}
    \label{fig:CW_fft}
    \addcontentsline{toc}{section}{Figure~\thefigure: Evolution of learned CW Raman-driving protocols and their Fourier spectra}
\end{figure*}

These span nearly two orders of magnitude (from $\sim\!0.06$~THz to $\sim\!4.38$~THz), providing both (i) \emph{slow} components near DC (the $\approx\!\pm 0.06$~THz terms) and in the sub-THz band ($\approx\!0.24$--$0.29$~THz) that act as envelopes or drift correctors over picosecond times, and (ii) \emph{faster} tones in the $1$--$4.4$~THz range that shape rapid excursions in the phase-space trajectory. Since time is represented in picoseconds, using THz (where $1~\mathrm{THz}=1/\mathrm{ps}$) ensures that the sinusoidal arguments $2\pi f_i t$ are dimensionless.
Panel (b) of Fig.~\ref{fig:ten_to_five} shows the five-frequency solution that matches the full model’s score while nulling the other five amplitudes. The retained frequencies (in THz) are
\[
\{f_i\}_{\text{kept}}\,[\mathrm{THz}] \approx 
\bigl\{\,2.825,\;2.695,\;0.2415,\;2.103,\;-0.2904\,\bigr\}.
\]
In particular, this subset preserves three mid-to-high THz tones (2.103, 2.695, 2.825~THz), which control the rapid oscillatory structure, together with two low-frequency components ($\approx\!0.24$ and $\approx\!0.29$~THz in magnitude) that supply slow envelopes for steering and stabilization. This suggests that the five-tone excitation suffices to reproduce the full dynamical performance while substantially reducing the experimental complexity.
Although Fig. \ref{fig:ten_to_five} demonstrates that the optimized CW protocol can be represented using only a small number of Fourier components, it does not reveal how such control strategies emerge during optimization. To gain further insight into the learning process, Fig. \ref{fig:CW_fft} shows representative protocols and their Fourier decompositions in different generations of the evolutionary search. Within 30 generations, the optimization process finds protocols capable of crossing the saddle point. In this particular example, by generation 62, the protocol is fully optimized, displaying a non-trivial $A(t)$ that induces the phase transition over a fully symmetric orbit. The Fourier decompositions of the successful protocols in Figure \ref{fig:CW_fft}b show five main frequency components of frequencies \textbf{lower} than the harmonic phonon frequency ($\omega_\textrm{ph}$), $\omega$ at 2.82THz ($0.93\omega_\textrm{ph}$), 2.69THz ($0.89\omega_\textrm{ph}$), 0.24THz ($0.08\omega_\textrm{ph}$), 2.10THz ($0.69\omega_\textrm{ph}$) and 0.29THz ($0.09\omega_\textrm{ph}$). The slightly detuned high-frequency oscillations produce an effective slow modulation, analogous to frequency mixing in dyadic lasers, that enables the $\simeq \omega_\textrm{ph}$ component of $A(t)$ to vanish near bifurcation.

\section{Additional details of protocol scoring}
\label{sec:score}
The score used to evaluate the optimized protocols is introduced in the main text. Here, we provide the complete mathematical formulation and additional details of its numerical evaluation. To quantify how well the trajectory of a dynamical system is stabilized around the saddle point, we analyze the trajectory in phase space, by dividing the total simulation time \( T \) into multiple time periods, as defined by the subsequent crossing \( t_i \) and \( t_{i+1} \),of the velocity \( \dot{x} = 0 \) axis. The trajectory of the system is divided into  \( N_{\text{orbit}} == \frac{T}{\Delta T_i}\)
where \( \Delta T_i = t_{i+1} - t_i \) is the duration of each time period.
The full trajectory \( x(t) \) of the system can then be expressed as:
\[
x(t) = \sum_{i=1}^{N_{\text{orbit}}} \theta_i(t)
\]
where
\[
\theta_i(t) = 
\begin{cases} 
x_i(t), & \text{for } t \in [t_i, t_{i+1}] \\
0, & \text{otherwise}
\end{cases}
\]
Here, \( x_i(t) \) represents the system’s trajectory within the \( i \)-th phase orbit. This approach allows the system’s overall behavior to be broken down into distinct intervals, facilitating a detailed analysis of the dynamics within each phase orbit. 
To further analyze the system’s behavior, the average position \( \langle x \rangle_i \) is calculated within each time period \( T_i \) as:
\[
\langle x \rangle_i = \frac{1}{\Delta T_i} \int_{t_i}^{t_{i+1}} x_i(t) \, dt
\]
The score for a given protocol is then determined as the maximum value of the symmetrized and normalized average position over all time periods \( T_i \), calculated as:

\[
\text{Score} = \max\left( 1 - \frac{|\langle x \rangle_i - x_{saddle}|}{\Delta x_{\text{ms}}} \right)
\]

and, \(\Delta x_{\text{ms}} \) is the distance between the minimum energy point, \( x_{\text{min}} \) and the saddle point, \( x_{\text{saddle}} \).
This score function quantifies the maximum deviation of the system’s average position from the saddle point in all phase orbits, providing a measure of the system’s dynamic behavior over time.
A score of 1 corresponds to a stable phase orbit where the average position \( \langle x \rangle \) consistently aligns with the saddle point (\( \langle x \rangle = 0 \)). This indicates that within each time period, the phase orbit remains centered around the saddle point, stabilizing the system at a critical phase transition point, and optimization is complete. Even if there is only one stable phase orbit that occurs within only one time period of the entire trajectory, the score is still 1, reflecting perfect stabilization at that critical point. A score of 0 indicates that the phase orbit is far from the saddle point and the average position \( \langle x \rangle \) significantly deviates from zero. 

\section{\label{sec:hyp_test}Hyperparameter testing}

At each generation, every neural network in the population generates a complete driving protocol from its current parameters. Each protocol is evaluated by integrating the corresponding equations of motion and computing its score (or ensemble fitness). The highest-scoring protocols are retained as elite members and selected as parents for the next generation, while new offspring are generated by applying Gaussian mutations to the parent network parameters. This evolutionary cycle is repeated until convergence or until the prescribed maximum number of generations is reached.

\begin{figure*}[htbp]
    \centering
    \includegraphics[width=0.8\textwidth]{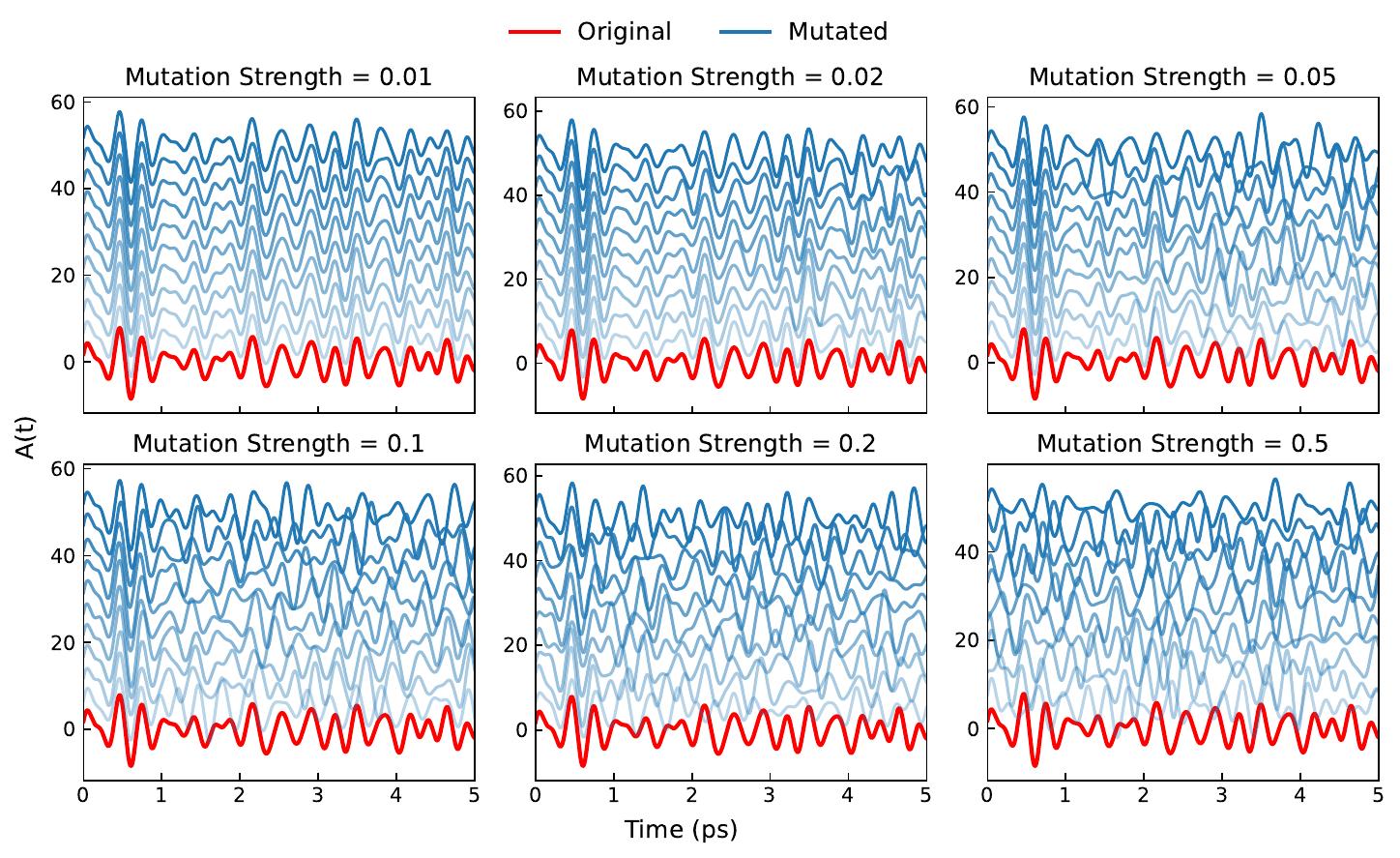}
    \caption{ Dependence of time-dependent continuous wave protocol on mutational strength}
    \label{fig:ms_fnn}
    \phantomsection
    \addcontentsline{toc}{section}{Figure~\ref{fig:ms_fnn}-~\ref{fig:ms_pulse}: Dependence of time-dependent protocols on mutational strength}
\end{figure*}

\begin{figure*}[htbp]
    \centering
    \includegraphics[width=0.8\textwidth]{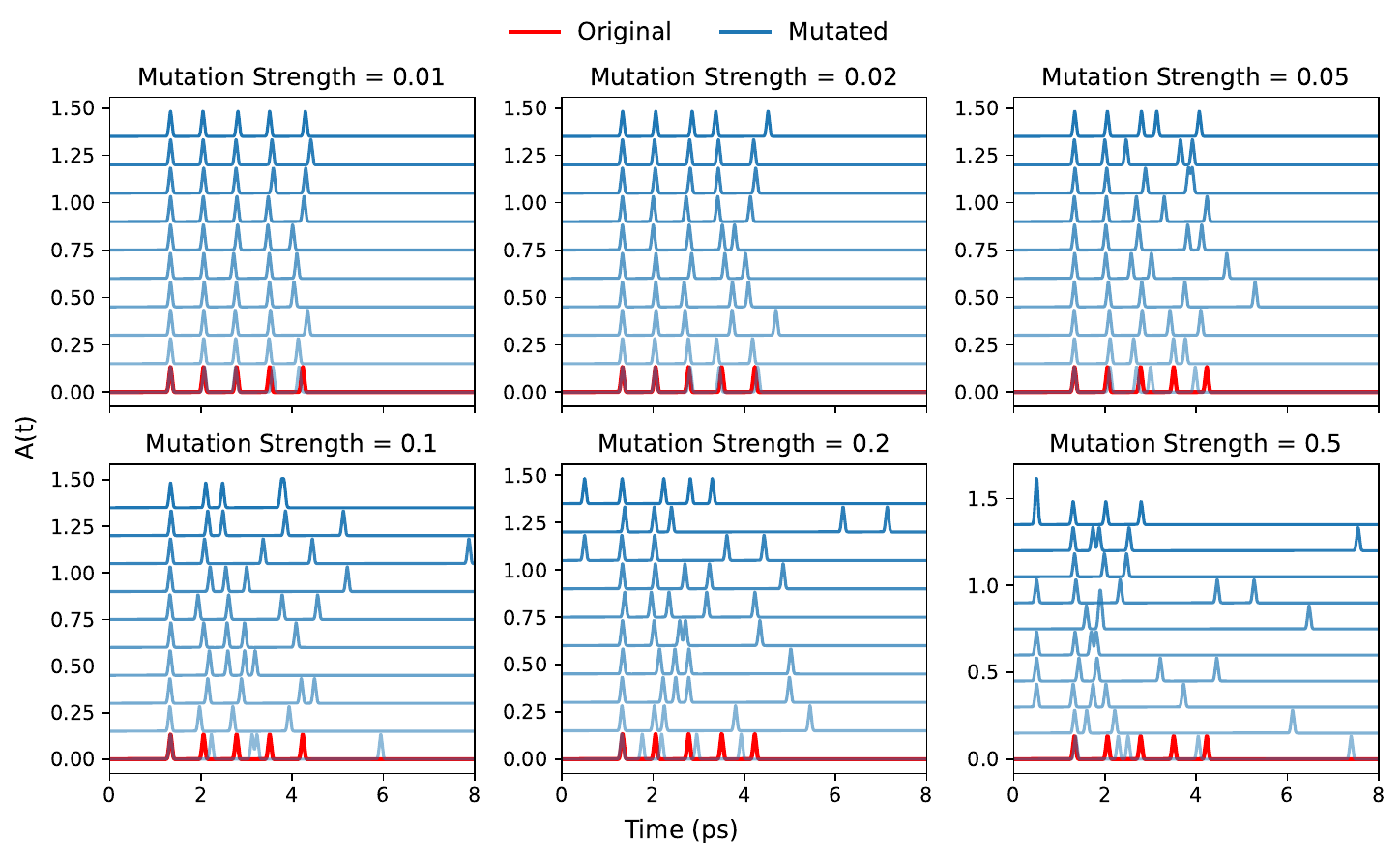}
    \caption{ Dependence of time-dependent pulsed protocols on mutational strength}
    \label{fig:ms_pulse}
\end{figure*}

\begin{figure*}[htbp]
    \centering
    \includegraphics[width=0.8\textwidth]{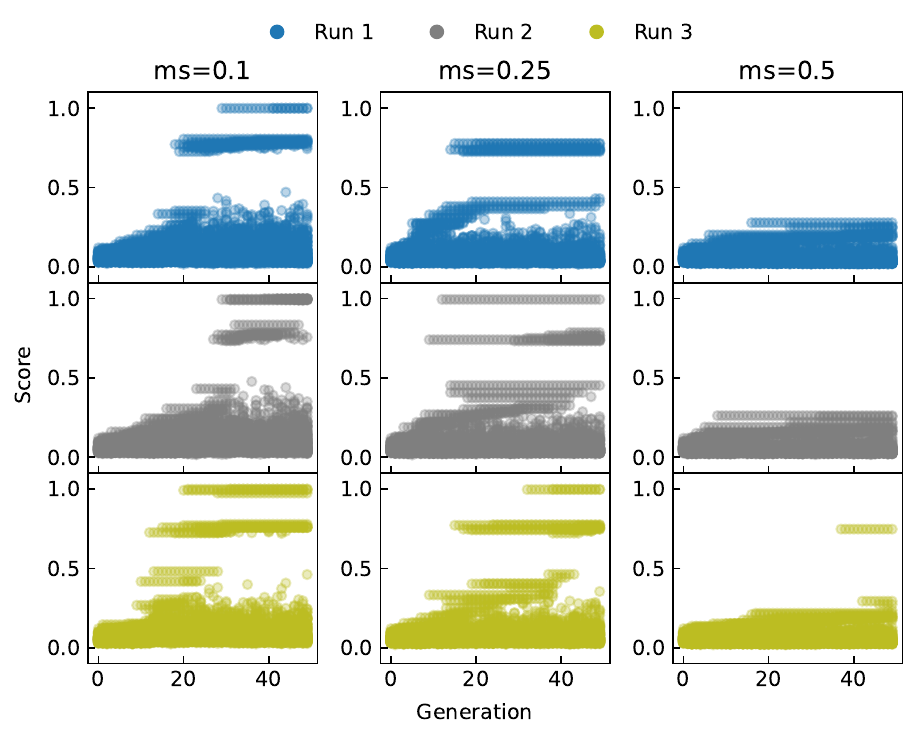}
    \caption{ Dependence of score of time-dependent protocols on mutation strength }
    \label{fig:hyp_ms}
    \phantomsection
    \addcontentsline{toc}{section}{Figure~\ref{fig:hyp_ms}-~\ref{fig:time_length}: Dependence of protocols score on various hyperparameters}
\end{figure*}

\begin{figure*}[htbp]
    \centering
    \includegraphics[width=0.8\textwidth]{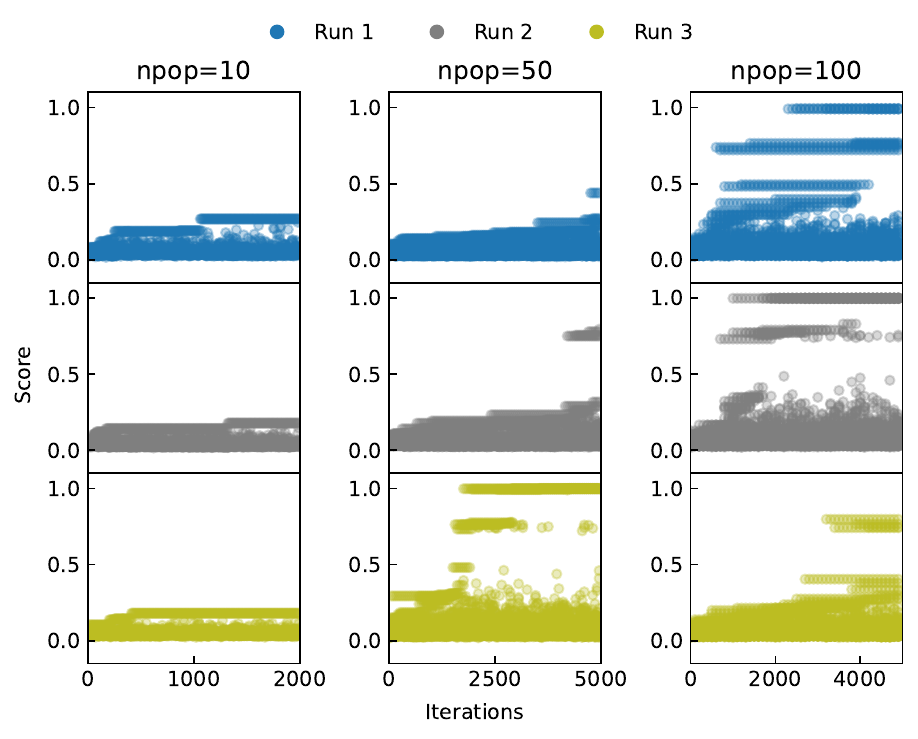}
    \caption{ Dependence of score of time-dependent protocols on number of population in each generation}
    \label{fig:hyp_npop}
\end{figure*}

\begin{figure*}[htbp]
    \centering
    \includegraphics[width=0.8\textwidth]{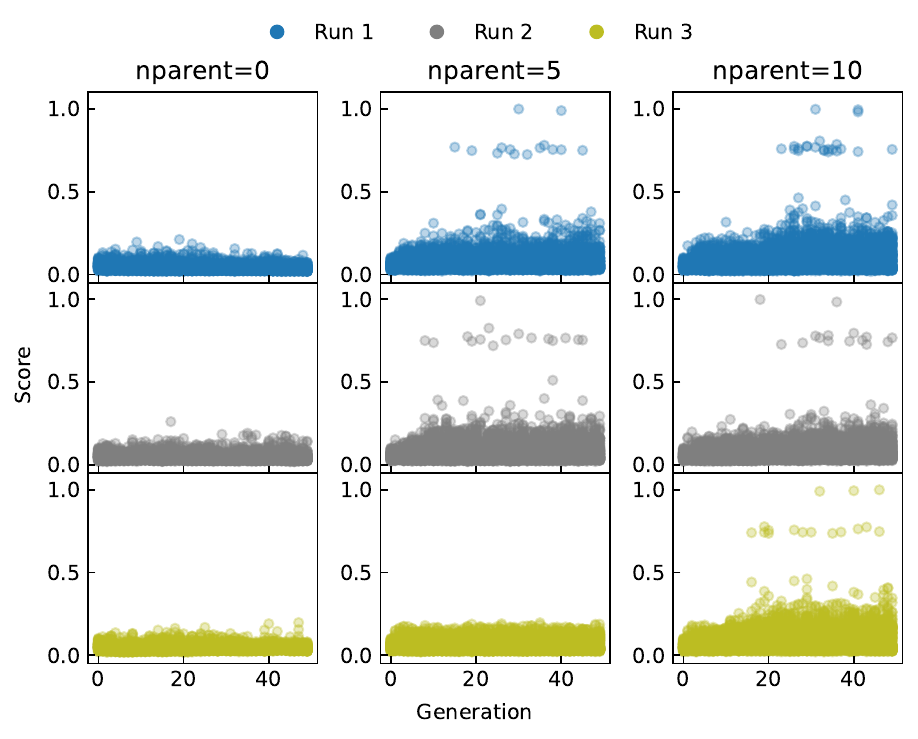}
    \caption{ Dependence of score of mutated time-dependent protocols on number of parent network in each generation}
    \label{fig:hyp_num_parent}
\end{figure*}

\begin{figure*}[htbp]
    \centering
    \includegraphics[width=4in]{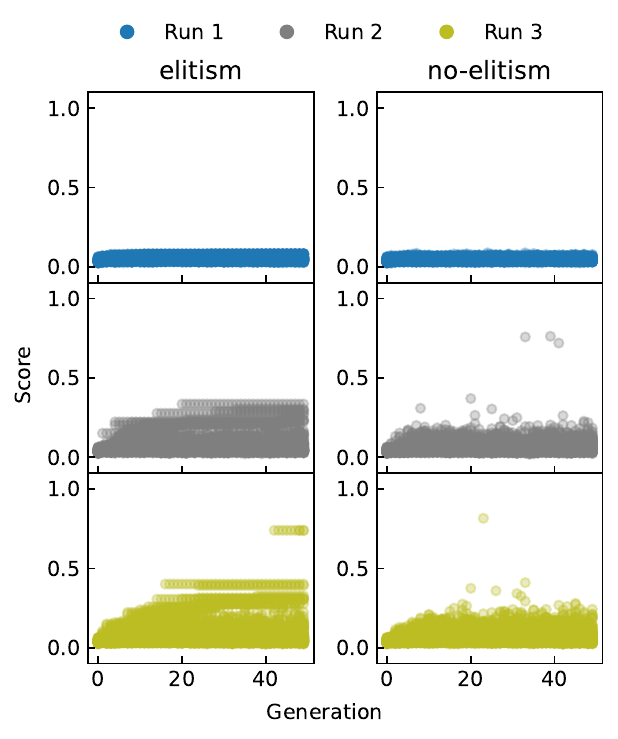}
    \caption{ Dependence of score of time-dependent protocols on inclusion of elitism at a mutation strength of 0.1}
    \label{fig:hyp_elit}
\end{figure*}

\begin{figure*}[htbp]
    \centering
    \includegraphics[width=0.8\textwidth]{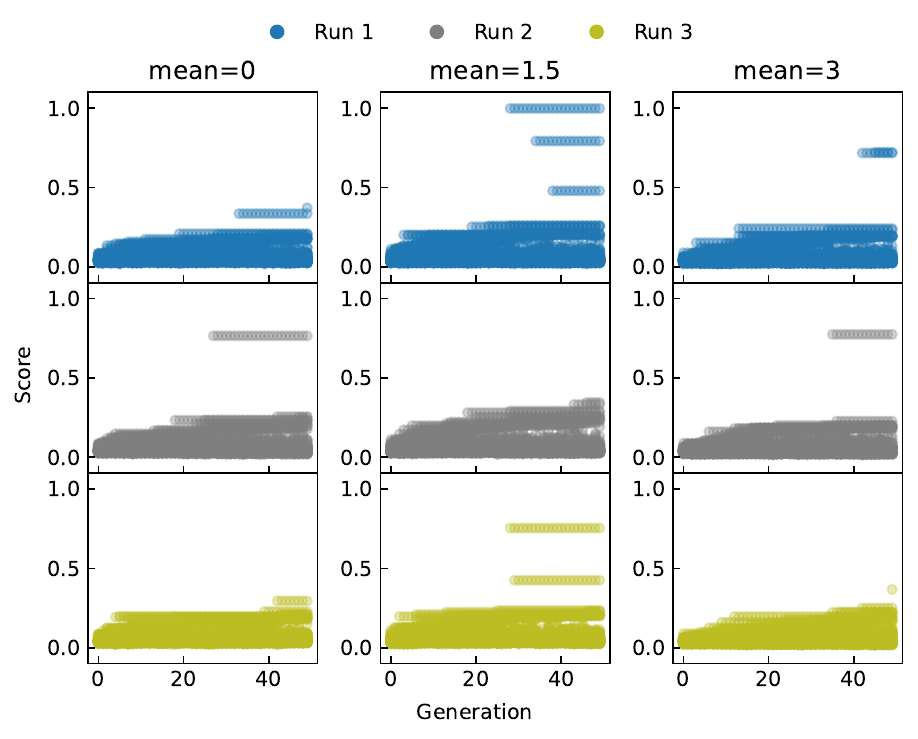}
    \caption{ Dependence of score of time-dependent protocols on random initial weights of FNN}
    \label{fig:hyp_init_wt}
\end{figure*}

\begin{figure*}[htbp]
    \centering
    \includegraphics[width=0.8\textwidth]{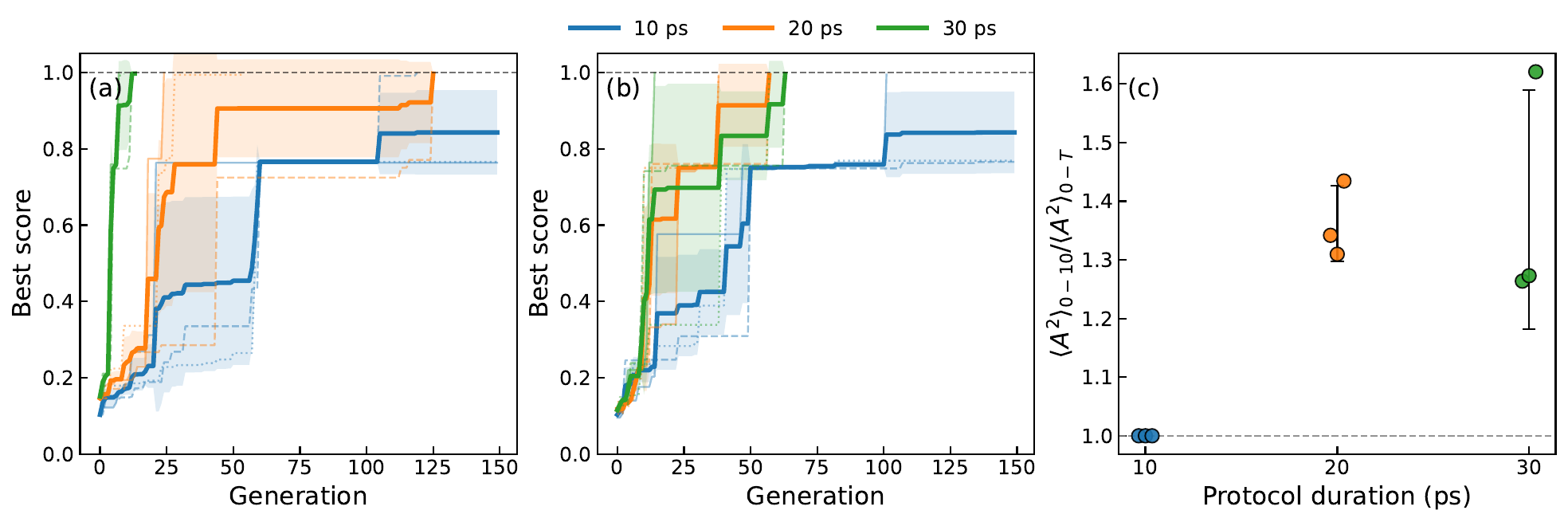}
    \caption{ (a) Best score versus generation for 10, 20, and 30 ps protocols when scoring is performed over the full protocol duration. (b) Best score versus generation for the control case where scoring is restricted to windows ending within the first 10 ps. (c) Ratio of the average intensity in the scored 0–10 ps window to the full-duration average intensity for the optimized protocols.}
    \label{fig:time_length}
\end{figure*}

We tested the sensitivity of the optimization to mutation strength, population size, number of parents, elitism, random initial weights, and total protocol duration. Figures~\ref{fig:ms_fnn}--\ref{fig:time_length} summarize the corresponding tests.

\textit{Mutation strength:} Figures~\ref{fig:ms_fnn} and~\ref{fig:ms_pulse} illustrate the effect of mutation strength on the generated protocols. Increasing the mutation strength produces progressively larger deviations from the parent protocol, enabling broader exploration of the protocol space. Figure~\ref{fig:hyp_ms} shows the corresponding effect on optimization, with larger mutation strengths producing greater variability between generations.

\textit{Population size:} Figure~\ref{fig:hyp_npop} shows the dependence of the optimization on population size. A population of $n_{\mathrm{pop}}=10$ does not achieve convergence, whereas all three runs with $n_{\mathrm{pop}}=100$ reach the optimal solution. For $n_{\mathrm{pop}}=50$, one of the three runs reaches convergence. Larger populations therefore improve exploration of the protocol space at the cost of increased computational expense.

\textit{Number of parents and elitism:} Figure~\ref{fig:hyp_num_parent} compares optimizations using $n_{\mathrm{parent}}=0$, 5, and 10. Using parent networks directs the search toward previously identified high-performing regions of parameter space, whereas $n_{\mathrm{parent}}=0$ corresponds to a less directed random search. Figure~\ref{fig:hyp_elit} shows that retaining elite protocols between generations leads to more consistent convergence by preserving high-scoring solutions.

\textit{Random initial weights:} Figure~\ref{fig:hyp_init_wt} compares initial weight distributions with means of 0, 1.5, and 3. The choice of initialization affects the regions of parameter space sampled during the early generations and consequently influences the rate at which high-scoring protocols are identified.

\textit{Protocol duration:} We also tested the dependence of the optimization on the total protocol duration using protocols of 10, 20, and 30~ps (Fig.~\ref{fig:time_length}). The protocols were compared using fixed time-averaged intensity normalization,
\begin{equation}
\langle A^2\rangle_{0-T}
=
\frac{1}{T}\int_0^T A^2(t)\,dt ,
\label{eq:time_avg_intensity}
\end{equation}
where $T$ is the protocol duration. Under this constraint, the total integrated drive is
\begin{equation}
\int_0^T A^2(t)\,dt
=
T\langle A^2\rangle_{0-T}.
\label{eq:integrated_drive}
\end{equation}
Consequently, the 20 and 30~ps protocols optimize more readily than the 10~ps protocol when the score is evaluated over the full protocol duration.

To determine whether this improvement results only from the availability of additional late-time scoring windows, we performed a control calculation in which trajectories were propagated for the full protocol duration but the score was evaluated only using windows ending within the first 10~ps. The longer protocols remain favorable under this restriction, indicating that their improved performance does not arise solely from additional scoring time. Instead, under fixed time-averaged intensity, longer protocols can redistribute intensity into the early scoring window. This redistribution is characterized by
\begin{equation}
\frac{\langle A^2\rangle_{0-10}}
     {\langle A^2\rangle_{0-T}},
\label{eq:intensity_redistribution}
\end{equation}
which is greater than unity for the optimized longer-duration protocols. We therefore use a 20~ps protocol as an intermediate choice between the shorter temporal window of the 10~ps protocol and the greater exposure and temporal redistribution permitted by the 30~ps protocol.

\section{Optical pulse parameters in physical units}
\label{sec:physical_units}

The conversion from the dimensionless field amplitudes used in the equations of motion to physical electric fields follows the internal unit system and
field--intensity relations described in Ref.~\cite{Haldar24}. In that
convention, the relevant derived units are
\begin{equation}
t_0 = 0.157461~{\rm ps},
\qquad
\mathcal{E}_0 = 4.180159\times10^7~{\rm V/m}
\end{equation}
The optical intensity is related to the electric field by
\begin{equation}
I(t)=\frac{|\mathcal{E}(t)|^2}{2\eta_0},
\qquad
\eta_0\simeq377~\Omega .
\end{equation}

For pulsed protocols, the raw Fourier-neural-network pulse train $A(t)$ is dimensionless. Before entering the equations of motion, it is converted to the normalized dimensionless electric-field amplitude
\begin{equation}
a_{\rm eff}^2(t)
=
s\frac{A^2(t)}
{\int A^2(t)\,dt_{\rm int}},
\label{eq:pulse_norm_si}
\end{equation}
where $t_{\rm int}$ is the time in internal units. The dimensionless scale factor
$s$ fixes the integrated normalized intensity,
$\int a_{\rm eff}^2(t)\,dt_{\rm int}=s$. We use $s=\alpha=1.55$ for the
impulsive Raman pulsed optimization and
$s=\mathrm{pulse\_area\_scale}=0.675$ for the displacive excitation
optimization.

The physical electric field is
\begin{equation}
\mathcal{E}(t)=a_{\rm eff}(t)\mathcal{E}_0 .
\end{equation}
The pulse fluence, obtained by integrating the optical intensity over time, is therefore proportional to $s$:
\begin{equation}
F
=
\int I(t)\,dt_{\rm SI}
=
\frac{\mathcal{E}_0^2t_0}{2\eta_0}s
=
36.52\,s~\mu{\rm J/cm^2}.
\label{eq:fluence_si}
\end{equation}
This gives fixed fluences of approximately
$57~\mu{\rm J/cm^2}$ for the impulsive Raman pulsed protocols and
$25~\mu{\rm J/cm^2}$ for the displacive protocols. For a circular pump spot
with diameter $400~\mu{\rm m}$, these correspond to total pulse energies of
approximately $0.07~\mu{\rm J}$ and $0.03~\mu{\rm J}$, respectively.

Although the fluence is fixed within each optimization, the peak electric field can vary because it depends on the maximum value of the normalized pulse envelope. For the impulsive Raman pulsed protocols, the peak electric field is approximately $40$--$60~{\rm MV/m}$ across generations. For the displacive protocols, the peak electric field remains approximately $26~{\rm MV/m}$. In the impulsive Raman pulsed one-dimensional optimization, the pulse separation $\Delta t$ was not constrained, so later-generation protocols can bring pulses closer together and increase the peak field through temporal overlap. In contrast, pulse overlap was not allowed in the displacive optimization to make the control problem more constrained; the five pulses therefore remain separated, and the peak electric field remains essentially unchanged across generations.

\section{\label{sec:efficiency}Determining driving force efficiency}
For protocols of five pulses (as used in this work) with a given total intensity
$\alpha = \int_{0}^{t} |A(t)|^2 dt$, for $R_{\textrm{max}}=\textrm{max}(|R(x(t)|))$ the driving force efficiency, $\eta$ can be approximated by the following:

\begin{equation}
    \eta = \frac{\int_{0}^{t} \text{sign} \left[ - \dot{x} R(x) \right] |A(t)|^2 |R(x)| dt}{\frac{1}{5} \alpha (4R_{\textrm{max}}  +  | R(x_{\text{min}})|)},
\label{Eq:eta}
\end{equation}
where $x$ and $\dot{x}$ stand for $x(t)$ and $\dot{x}(t)$, respectively.

\begin{figure*}[htbp]
    \centering
    \includegraphics[width=0.8\textwidth]{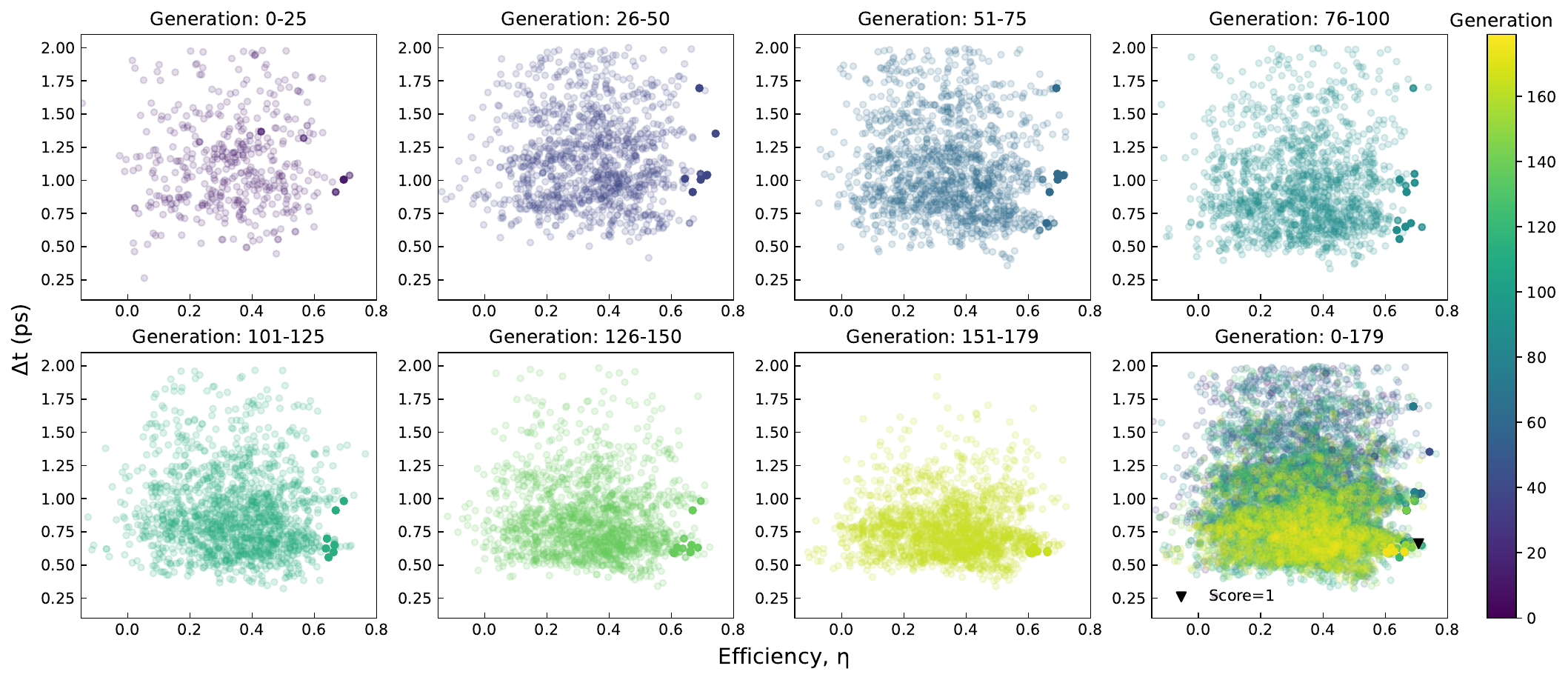}
    \caption{Distribution of Protocols Across Generations. This figure shows the distribution of protocol efficiency (\(\eta\)) and pulse timing spread (\(\Delta t\)) across generations. Later generations have more protocols converging toward high efficiency and low dissipation, reflecting improved optimization.}
    \label{fig:eff_groups}
    \phantomsection
    \addcontentsline{toc}{section}{Figure~\thefigure: Distribution of Protocols Across Generations.}
\end{figure*}

\begin{figure*}[htbp]
    \centering
    \includegraphics[width=0.8\textwidth]{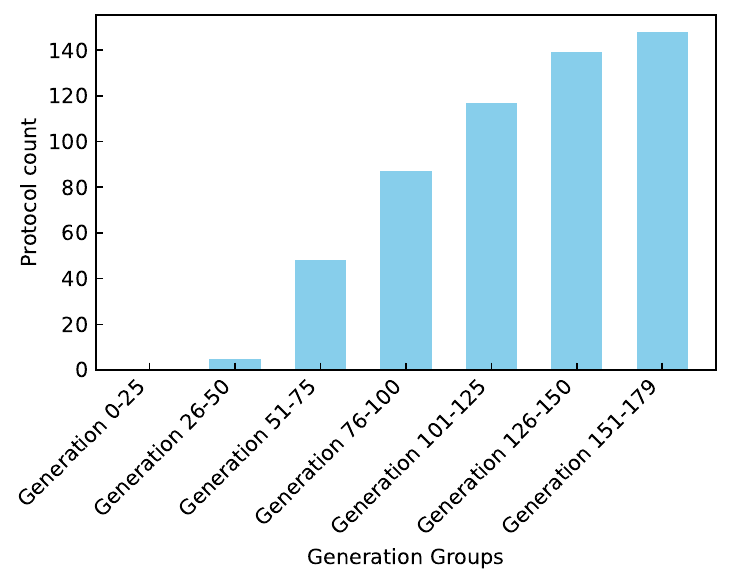}
    \caption{Number of Near-Optimal Protocols Per Generation. Bar plots display the increase in the number of protocols achieving near-optimal \(\Delta t\) and \(\eta\) values across generations, with later generations showing more convergence towards optimal solutions.}
    \label{fig:bar_groups}
    \phantomsection
    \addcontentsline{toc}{section}{Figure~\thefigure: Number of Near-Optimal Protocols Per Generation}
\end{figure*}

\begin{figure*}[htbp]
    \centering
    \includegraphics[width=0.8\textwidth]{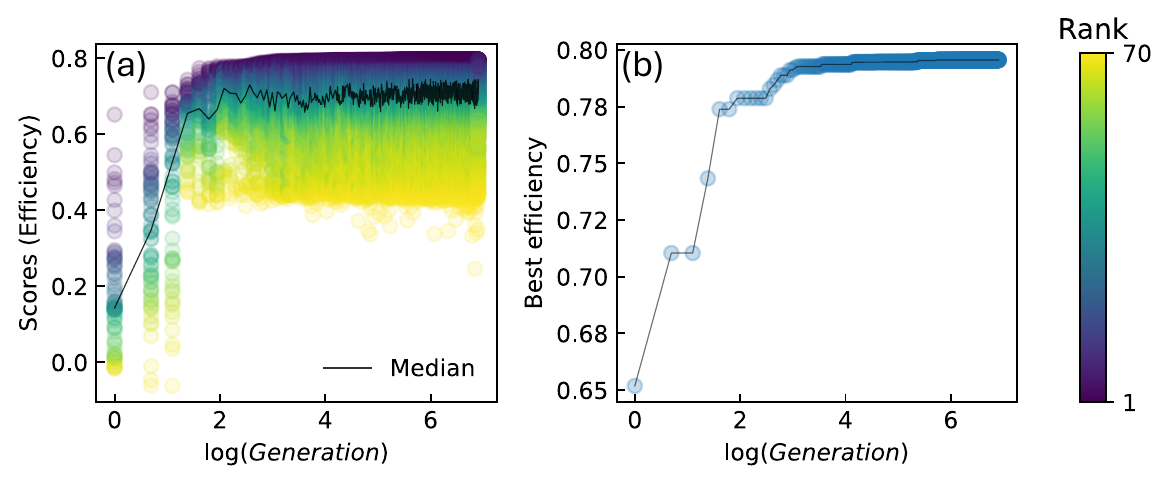}
    \caption{(a)Efficiency as scores across generations as the neural network optimizes protocols and (b)Best efficiency achieved at each generation. The scores steadily increase, indicating the network's learning progress, and begin to stabilize as it approaches the efficiency limit, demonstrating how the optimization protocol consistently identifies and refines the highest efficiency solutions}
    \label{fig:eff_scores}
    \phantomsection
    \addcontentsline{toc}{section}{Figure~\thefigure: Efficiency as scores across generations}
\end{figure*}

 The learning process for optimizing control strategies for pulsed 1D driving can be characterized by looking at the qualitative descriptors that govern efficiency: the overall pulse time spread \(\Delta t\) (as low as possible to minimize damping and dissipation) and the efficiency of driving force \(\eta\), quantifying how close the system was to its maximum of optomechanical coupling when irradiated. We note that a perfect efficiency of \(\eta = 1\) would only be achievable for infinitely brief pulses if the pulses coincided with the extrema of \(R(x)\). For early generations, a wider distribution of protocols is observed for both \(\Delta t\) and \(\eta\). As optimization progresses, a growing proportion of protocols increasingly converge to a region characterized by both high efficiency and minimal \(\Delta t\), as shown in Figures~\ref{fig:eff_groups} and~\ref{fig:bar_groups}. Figure~\ref{fig:eff_scores} shows that the algorithm can also maximize \(\eta\) if used as the only optimization metric. The efficiency approaches a practical limit of approximately 0.8, highlighting the network's convergence toward optimal performance.
 
\begin{figure*}[htbp]
    \centering
    \includegraphics[width=\textwidth]{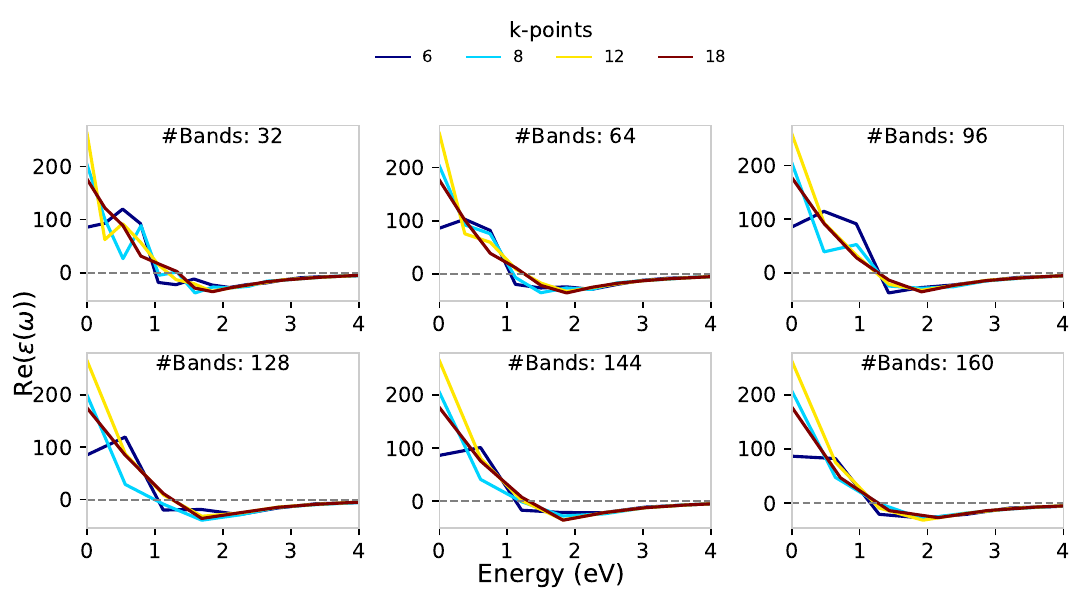}
    \caption{ Density-density real part of the dielectric function for different number of k-points (number, N, in the legend refers to the NxNxN grid) and bands for ground state configuration, zoomed in over lower frequency range}
    \label{fig:conv_gs_lowx}
    \phantomsection
    \addcontentsline{toc}{section}{Figures \ref{fig:conv_gs_lowx}-\ref{fig:conv_saddle_lowy}: k-point and number of band convergence for dielectric function}
\end{figure*}

\begin{figure*}[htbp]
    \centering
    \includegraphics[width=\textwidth]{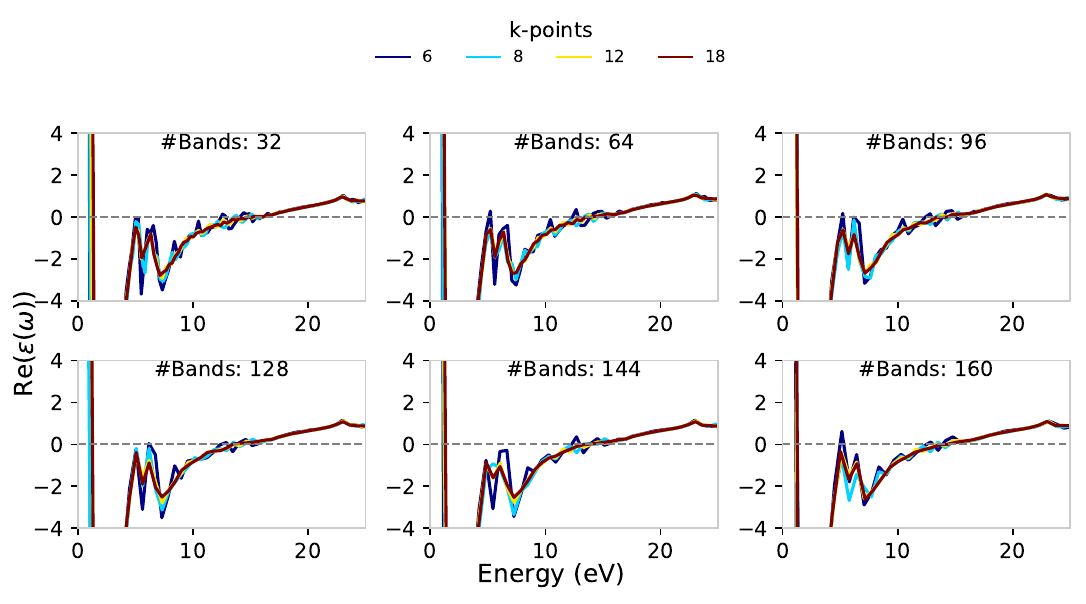}
    \caption{ Density-density real part of the dielectric function for different number of k-points (number, N, in the legend refers to the NxNxN grid) and bands for ground state configuration, zoomed in over lower dielectric value range}
    \label{fig:conv_gs_lowy}
\end{figure*}


\begin{figure*}[htbp]
    \centering
    \includegraphics[width=\textwidth]{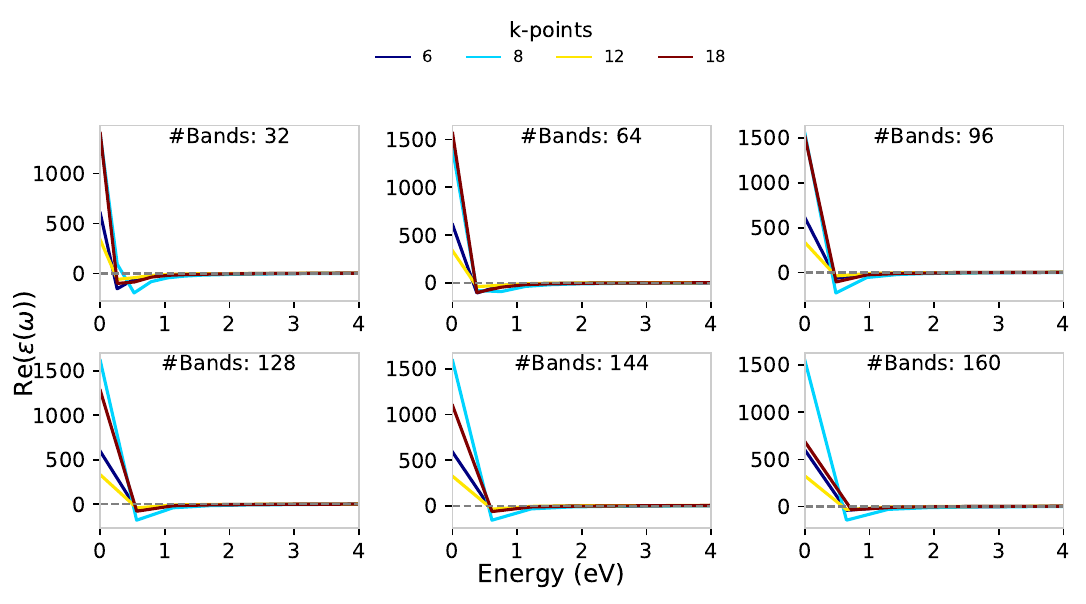}
    \caption{ Density-density real part of the dielectric function for different number of k-points (number, N, in the legend refers to the NxNxN grid) and bands for saddle point configuration, zoomed in over lower frequency range}
    \label{fig:conv_saddle_lowx}
\end{figure*}

\begin{figure*}[htbp]
    \centering
    \includegraphics[width=\textwidth]{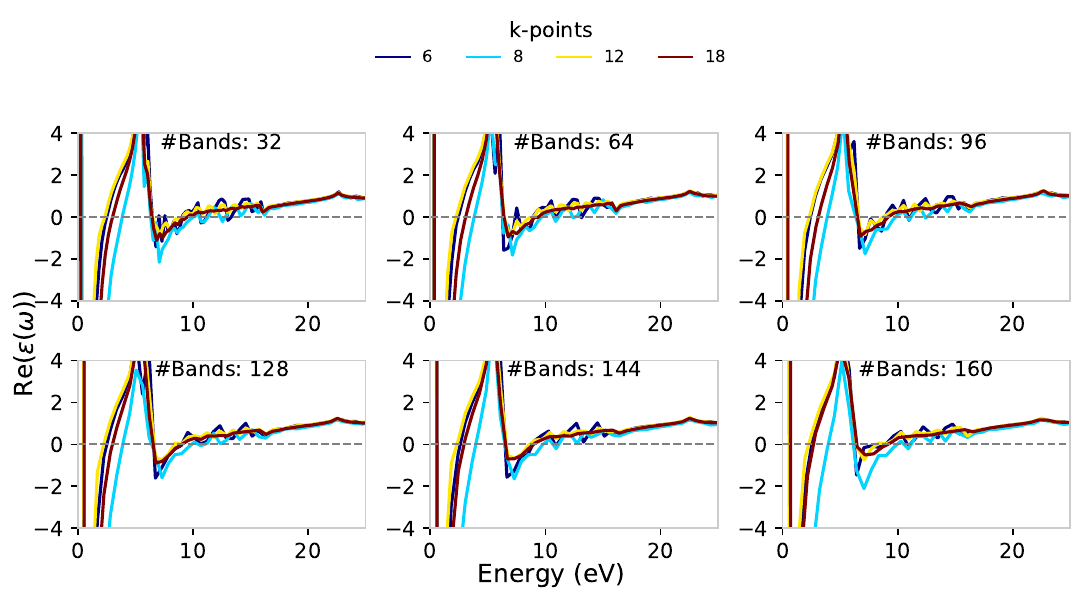}
    \caption{ Density-density real part of the dielectric function for different number of k-points (number, N, in the legend refers to the NxNxN grid) and bands for saddle point configuration, zoomed in over lower dielectric value range}
    \label{fig:conv_saddle_lowy}
\end{figure*}

\begin{figure*}[htbp]
    \centering
    \includegraphics[width=\textwidth]{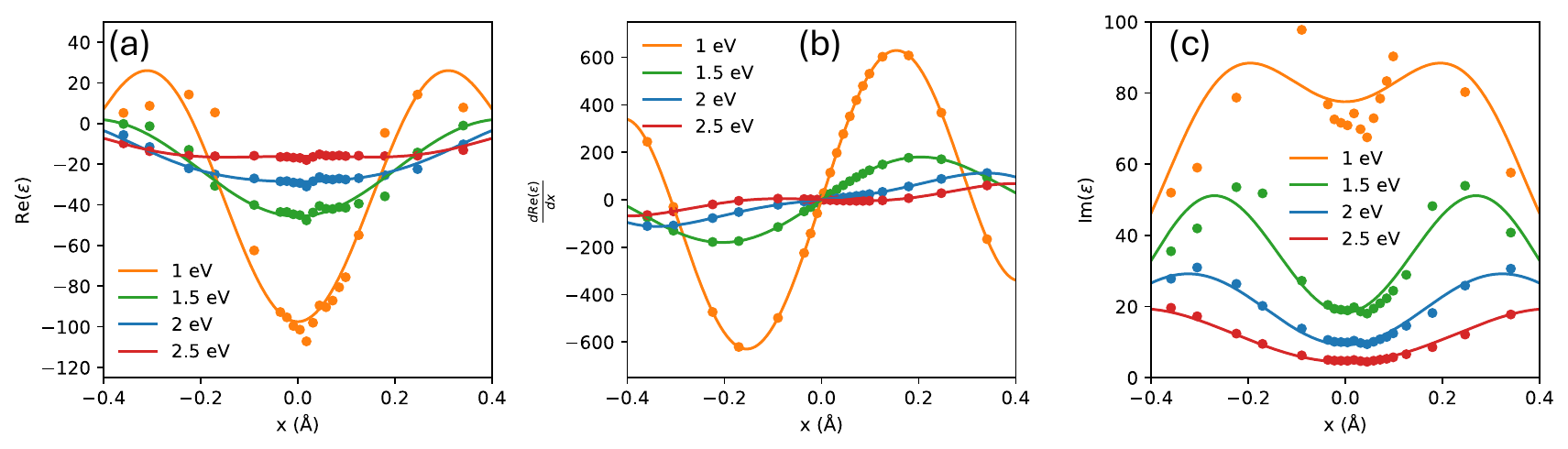}
    \caption{ (a) Real part of dielectric function plotted as function of phonon displacement coordinate; the more negative the value, the more metallic the behavior, (b) Raman cross-section and (c) Imaginary part of dielectric function at four different frequencies. }
    \label{fig:Rx_omega}
    \phantomsection
    \addcontentsline{toc}{section}{Figure~\thefigure: Dielectric function and Raman cross-section at four different frequencies}
\end{figure*}
\begin{figure*}[htbp]
    \centering
    \includegraphics[width=\textwidth]{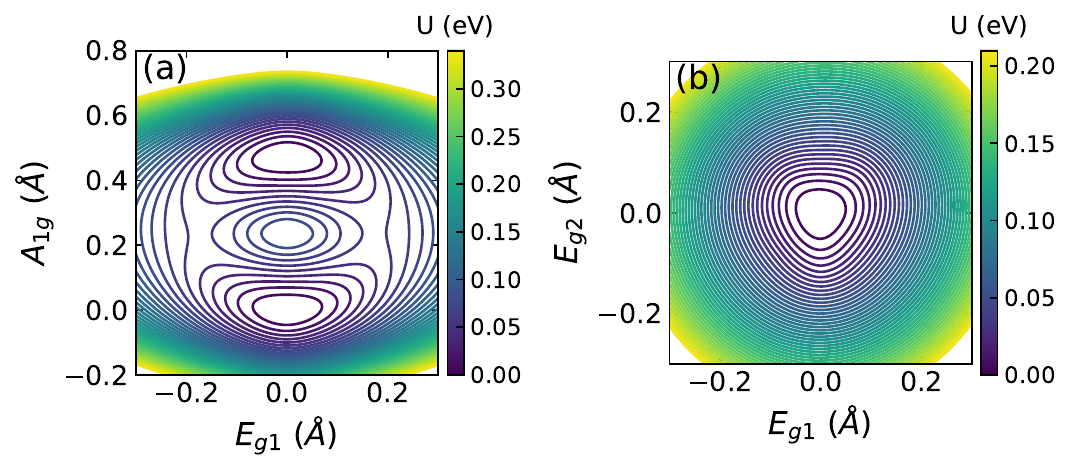}
    \caption{
    Two-dimensional projections of the multidimensional DFT potential energy surface used for the CW Raman-driving optimization on the three-dimensional PES. The energy surface was constructed from DFT calculations on Sobol-sampled configurations and interpolated using a radial-basis-function (RBF). (a) Projection onto the \(A_{1g}\)-\(E_{g1}\) plane with \(E_{g2}=0\). (b) Projection onto the \(E_{g2}\)-\(E_{g1}\) plane with \(A_{1g}=0\). Contours show relative energies with respect to the global minimum.
    }
    \label{fig:3d_pes}
    \phantomsection
    \addcontentsline{toc}{section}{Figure~\thefigure: Two-dimensional projections of the multidimensional DFT potential energy surface}
\end{figure*}

\clearpage

\begin{figure*}[htbp]
    \centering
    \includegraphics[width=0.8\textwidth]{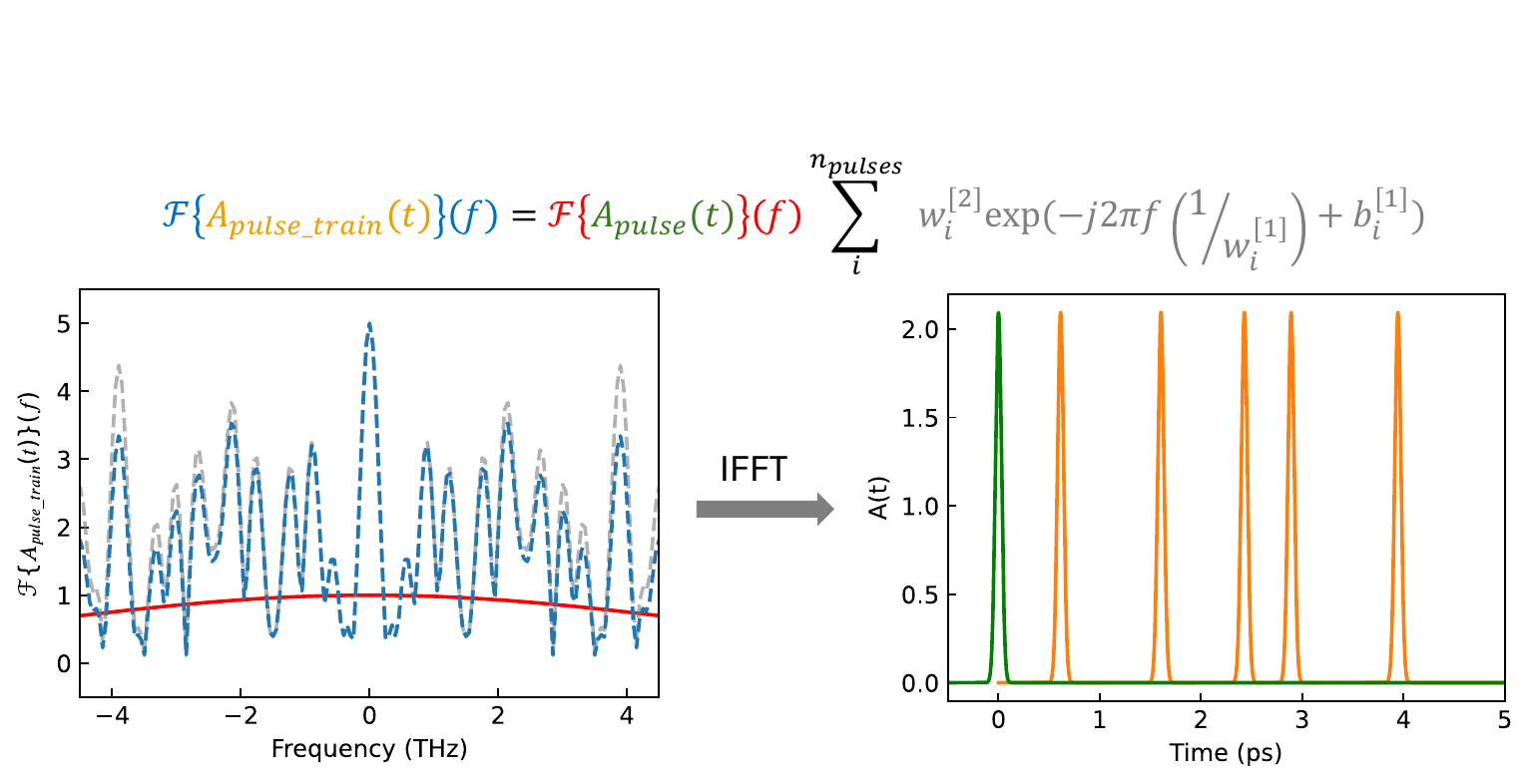}
    \caption{ Pulse driven protocol (right) obtained as the inverse FFT of the product of the two Fourier transform components: the term in the red is obtained experimentally and the term in gray is optimized using the same FNN architecture}
    \label{fig:ifft_pulse}
    \phantomsection
    \addcontentsline{toc}{section}{Figure~\thefigure: Pulsed protocol generation scheme}
\end{figure*}

\begin{figure*}[htbp]
    \centering
    \includegraphics[width=0.8\textwidth]{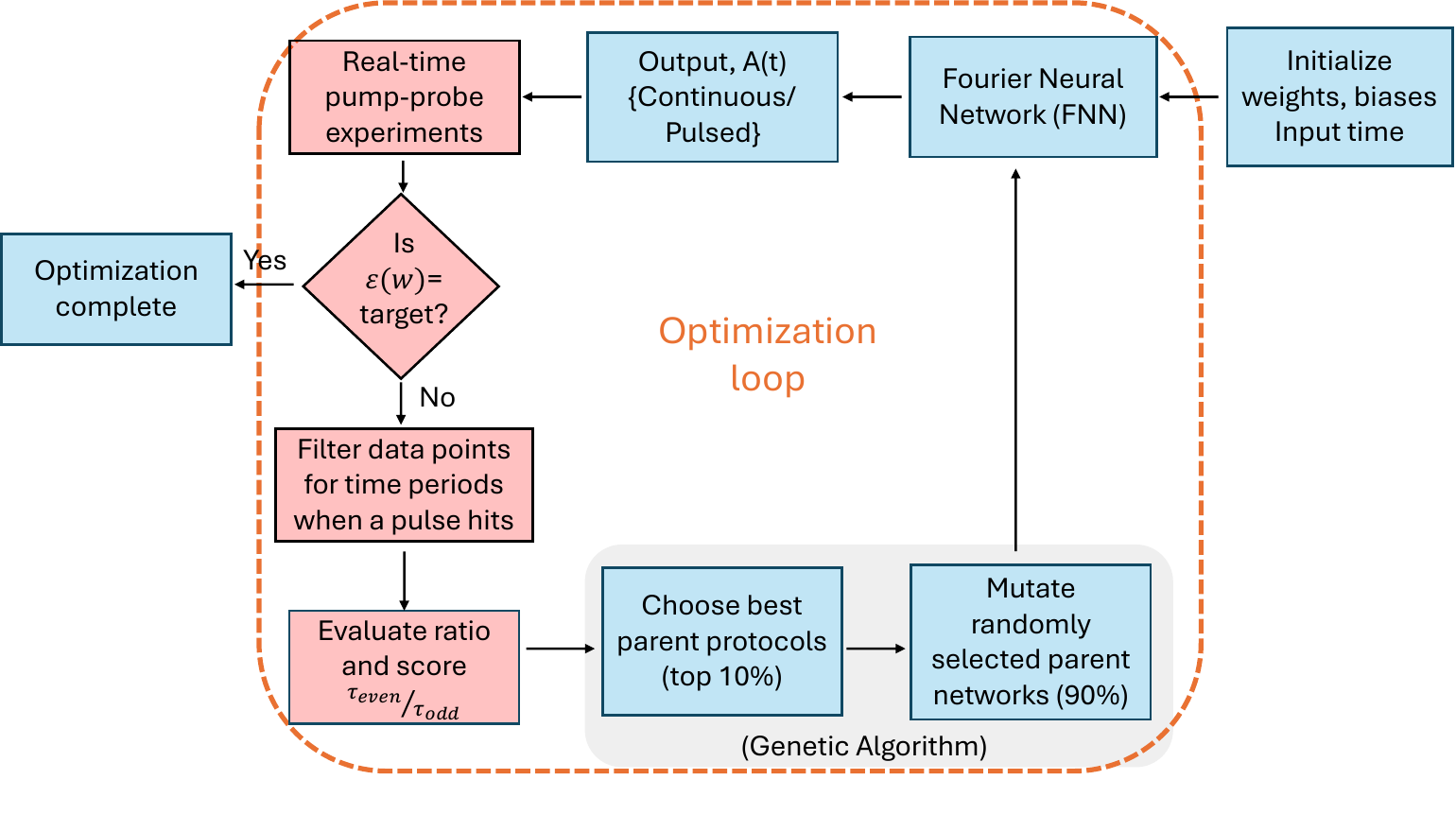}
    \caption{Optimization framework of time-dependent illumination protocols in an experimental setup}
    \label{fig:expt_opt}
    \phantomsection
    \addcontentsline{toc}{section}{Figure~\thefigure: Optimization framework of time-dependent protocols in an experimental setup}
\end{figure*}

\bibliography{biblio}